\documentclass[
  aps,
  prx,
  reprint,
  superscriptaddress,
  longbibliography,
  amsmath,amssymb,
  floatfix,
]{revtex4-2}

\usepackage[T1]{fontenc}
\usepackage[utf8]{inputenc}
\usepackage{bm}
\usepackage{braket}
\usepackage{graphicx}
\usepackage{xcolor}
\usepackage{hyperref}
\usepackage{subcaption}
\usepackage{ragged2e}
\usepackage[justification=justified,singlelinecheck=false]{caption}
\usepackage{amsmath}
\usepackage{svg}
\usepackage{caption}
\usepackage{xr-hyper}
\usepackage{hyperref}
\usepackage{xurl}
\usepackage{placeins}

\begin{document}

\title{Quantum Simulation of Two-Dimensional Free Dirac Hamiltonian in Multimode Circuit QED}

\author{Jiwon Kang}
\affiliation{Department of Physics, KAIST, Daejeon 34141, Republic of Korea}

\author{Jiuk Lee}
\affiliation{Department of Physics, KAIST, Daejeon 34141, Republic of Korea}

\author{Eliya Blumenthal}
\affiliation{Department of Physics, Technion - Israel Institute of Technology, Haifa 32000, Israel}

\author{Shay Hacohen-Gourgy}
\affiliation{Department of Physics, Technion - Israel Institute of Technology, Haifa 32000, Israel}

\author{Eunseong Kim}
\email{eunseong@kaist.edu}
\affiliation{Department of Physics, KAIST, Daejeon 34141, Republic of Korea}
\affiliation{The Graduate School of Quantum Science and Technology, KAIST, Daejeon 34141, Republic of Korea}

\date{\today}

\begin{abstract}
Two-dimensional massive Dirac systems feature a gapped Dirac-cone dispersion central to a broad range of phenomena in condensed-matter and topological physics. While quantum simulations have demonstrated one-dimensional Dirac dynamics and two-dimensional massless Weyl dynamics, programmable simulation of massive two-dimensional Dirac dynamics remains experimentally unexplored. Here, we realize a programmable two-dimensional free Dirac Hamiltonian in circuit QED, with independently tunable spin-momentum couplings and mass, using a single Rabi-driven qubit coupled to two modes of a multimode cavity. Using this Hamiltonian, we observe rotational Zitterbewegung of a two-dimensional massive Dirac particle and its dependence on the effective mass. Time-dependent master-equation simulations incorporating measured decoherence, corrections beyond the rotating-wave approximation (RWA), and anharmonicity of the transmon reproduce the observed dynamics. Our results establish multimode circuit QED as a compact, programmable platform for higher-dimensional relativistic quantum dynamics and provide a foundation for exploring dynamical and topological phenomena in gapped Dirac systems.

\end{abstract}

\maketitle

\section{Introduction}

The Dirac equation provides a fundamental framework for describing relativistic spin-1/2 particles by combining quantum mechanics with special relativity \cite{Thaller2013}. One of its paradigmatic predictions is Zitterbewegung, the oscillatory motion of a Dirac particle arising from interference between positive- and negative-energy states \cite{Schrodinger1930}. In two dimensions, a finite Dirac mass enriches this dynamics, giving the Zitterbewegung a rotational component with a well-defined handedness. Because Zitterbewegung occurs at extremely high frequencies and small amplitudes for free Dirac particles, controllable quantum systems have been developed to emulate Dirac Hamiltonians and probe the dynamics on experimentally accessible scales \cite{Feynman1982, Buluta2009}.

Quantum simulations with trapped ions have demonstrated one-dimensional Dirac dynamics and the associated Zitterbewegung \cite{Gerritsma2010, Lamata2007, Lamata2011, casanova2010}. A subsequent trapped-ion experiment extended these simulations to two dimensions by realizing a Weyl Hamiltonian and investigated the dynamics under effective magnetic fields \cite{Jiang2022}. However, the two-dimensional Zitterbewegung of free massive Dirac particles has not yet been realized in a quantum simulation. In two dimensions, introducing a finite Dirac mass opens a gap at the Dirac point and gives rise to finite local Berry curvature in momentum space \cite{Cayssol_2021, RevModPhys.82.1959}. The sign of this Berry curvature is associated with the handedness of the Zitterbewegung, linking the real-space dynamics of a massive Dirac particle to its underlying band geometry \cite{chiral_zb, zb_sym_broken}.

Several approaches have been made to simulate two-dimensional massive Dirac dynamics \cite{zb_sym_broken, Zhang2025, Liu:20,PhysRevLett.99.236809}. In photonic honeycomb lattices, sublattice-symmetry breaking opens a gap at the Dirac points, enabling massive Dirac dynamics to be reproduced through classical optical-wave propagation \cite{zb_sym_broken}. However, the effective Hamiltonian is largely encoded in the lattice structure, limiting independent control of its constituent terms. A superconducting-qubit experiment has also simulated massive Dirac dynamics and Zitterbewegung, but the momentum components were mapped onto qubit–qubit coupling strengths and thus treated as fixed parameters rather than dynamical quantum degrees of freedom \cite{Zhang2025}. These approaches differ from a fully programmable quantum simulator of two-dimensional massive Dirac dynamics, in which both momentum components are represented by quantum operators and the spinor–momentum couplings and effective Dirac mass can be independently controlled.

Here, we realize a tunable two-dimensional massive Dirac Hamiltonian using a single transmon coupled to two cavity modes \cite{PhysRevLett.127.107701}. Our approach exploits a Rabi-driven qubit to engineer an effective Dirac spinor degree of freedom, while the conjugate quadratures of each cavity mode represent the position and momentum operators along one spatial dimension \cite{wn7t-pyrq}. Sideband drives of the multimode cavity then provide controlled spinor–momentum couplings along the two spatial dimensions, while the detuning between the Rabi and double-sideband frequencies generates the effective mass term. This architecture enables a programmable quantum simulation of two-dimensional Dirac dynamics encompassing both spinor and spatial degrees of freedom. After benchmarking the simulation in one dimension, we implement the two-dimensional free Dirac Hamiltonian and observe its Zitterbewegung dynamics, finding that the rotational frequency increases while the amplitude decreases with increasing effective mass. We further compare the experimental dynamics with time-dependent master-equation simulations and identify spin-locking decoherence of the Rabi-driven qubit, corrections beyond the RWA, finite transmon anharmonicity, and control-parameter fluctuations as the main limitations of the simulation \cite{Lu2022, Bylander2011}.

\section{Circuit QED implementation}

\begin{figure*}[t]
    \centering
    \includegraphics[width = 0.99\textwidth]{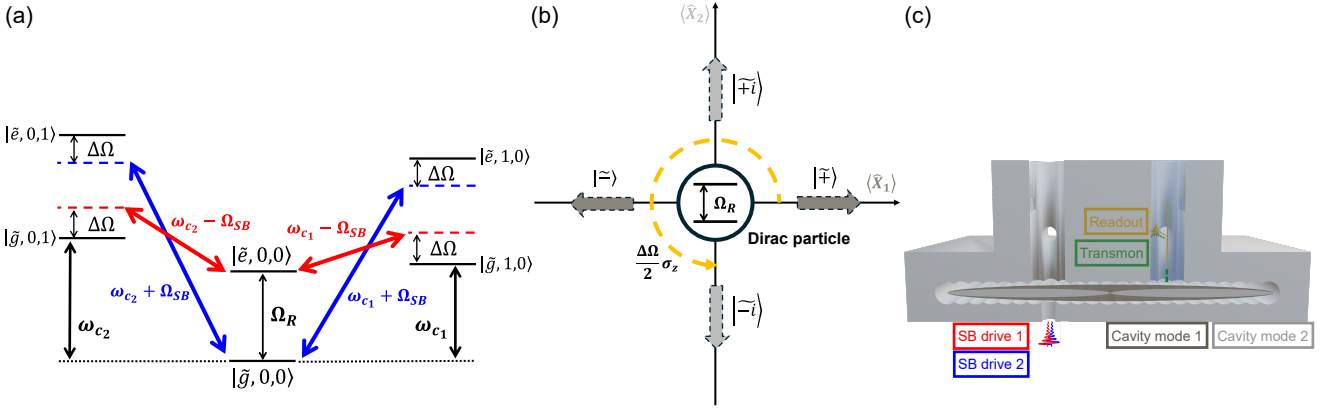}
    \caption{\justifying(a) Energy-level diagram of a Rabi-driven qubit interacting with cavity modes via double-sideband drives. (b) Schematic illustration of the dynamics governed by the effective Dirac Hamiltonian: $\sigma_x$ for $\delta_1=-\pi/2$ and $\sigma_y$ for $\delta_2=0$. The Rabi-driven effective qubit represents the Dirac spinor, and one of the phase-space quadratures of each cavity mode corresponds to the position operator of the simulated Dirac particle. The arrows denote the spin-dependent cavity displacement direction. (c) Schematic of the experimental setup comprising a flute cavity integrated with a transmon qubit and a microstrip readout resonator.}
    \label{fig1}
\end{figure*}

The target Hamiltonian is the free massive Dirac Hamiltonian in 2+1 dimensions, represented as
\begin{equation}
H_D
= c\hat{P}_1\sigma_x+c\hat{P}_2\sigma_y+mc^2\sigma_z,
\label{Dirac_Hamiltonian_Original}
\end{equation}
where $c$ is the speed of light, $\hat{P}_1$ and $\hat{P}_2$ are the momentum operators along the two spatial dimensions, and $m$ is the mass of Dirac particle \cite{Arrighi_2018, wn7t-pyrq}. The massless limit yields a gapless linear dispersion, whereas a finite mass opens an energy gap and realizes the massive Dirac regime explored in this work.

The principle for implementing the two-dimensional Dirac Hamiltonian of Eq.~\eqref{Dirac_Hamiltonian_Original} in circuit QED is illustrated schematically in Fig.~\ref{fig1}(a). In the rotating frame of the effective qubit, a Rabi drive with frequency $\Omega_R$ defines the dressed qubit basis, which serves as the Dirac spinor, with the dressed states $\ket{\tilde{g}}$ and $\ket{\tilde{e}}$ corresponding to the negative- and positive-energy eigenstates, respectively (corresponding to the $\ket{+}$ and $\ket{-}$ states in the lab frame, respectively). Double-sideband drives with equal amplitudes are applied to each cavity mode at frequencies detuned by $\pm \Omega_{\mathrm{SB}}$ from the cavity resonance, generating the required spinor-momentum coupling through interactions between the effective qubit basis states and the cavity photon-number states \cite{Hacohen2016}. In addition, a finite detuning between the Rabi frequency and the sideband frequency, $\Delta \Omega \equiv \Omega_R - \Omega_{\mathrm{SB}}$, produces an effective rotation of the qubit around the $z$-axis and sets the effective mass term of the simulated Dirac particle \cite{wn7t-pyrq}. 

Applying the rotating-frame transformations of the cavity frequencies, the Rabi drive, and the time-dependent cavity displacement, followed by the RWA, yields the effective Dirac Hamiltonian (\(n=1,2\)):
\begin{equation}
\frac{H'_{nD}}{\hbar}=\sum_{j=1}^{n}\left[-\frac{i\chi_j\alpha_j}{4}\left(\hat{a}_j-\hat{a}_j^\dagger \right)\sigma_{\delta_j+\frac{\pi}{2}}\right]+
\frac{\Delta \Omega}{2}\sigma_z.
\label{Dirac_Hamiltonian}
\end{equation}
Here, $\chi_j$ is the coupling strength between the qubit and the $j$-th cavity mode, $\alpha_j$ is the corresponding double-sideband amplitude, $\hat{a}_j$ ($\hat{a}_j^\dagger$) is the annihilation (creation) operator of the cavity modes. The parameter $\delta_j$ is an effective phase determined by the initial coherent state of the corresponding cavity mode.

The correspondence between Eq.~\eqref{Dirac_Hamiltonian} and the target Hamiltonian in Eq.~\eqref{Dirac_Hamiltonian_Original} can now be made explicit. The phase-space quadratures of each cavity mode are mapped onto the position and momentum operators as $\hat{X}_j=(\hat{a}_j+\hat{a}_j^\dagger)/2$ and $\hat{P}_j=i(\hat{a}_j^\dagger-\hat{a}_j)$, satisfying $[\hat{X}_j,\hat{P}_j]=i$. Furthermore, since $\delta_j$ determines the spinor-momentum coupling axis through $\sigma_{\delta_j}=\sigma_x\cos\delta_j+\sigma_y\sin\delta_j$, choosing $\delta_1=-\pi/2$ and $\delta_2=0$ yields $\sigma_{\delta_1+\pi/2}=\sigma_x$ and $\sigma_{\delta_2+\pi/2}=\sigma_y$, respectively. With these identifications, Eq.~\eqref{Dirac_Hamiltonian} reproduces the two-dimensional Dirac Hamiltonian in Eq.~\eqref{Dirac_Hamiltonian_Original}. A detailed derivation of the effective Hamiltonian and its mapping onto the Dirac Hamiltonian is provided in \cite{wn7t-pyrq}.

The dynamics of the Dirac particle governed by Eq.~\eqref{Dirac_Hamiltonian} are illustrated schematically in Fig.~\ref{fig1}(b). The first term generates a spin-dependent displacement of the cavity state, whose direction is determined by the instantaneous orientation of the effective Dirac spinor. More specifically, the velocity operators are directly related to the corresponding spin components through the Heisenberg equation of motion,
$\hat{v}_j \equiv \dot{\hat{X}}_j
= \frac{i}{\hbar}[H'_{nD},\hat{X}_j]
= \frac{\chi_j\alpha_j}{4} \sigma_{\delta_j+\frac{\pi}{2}}$. The second term rotates the effective Dirac spinor about the $z$ axis, continuously changing the spin-dependent displacement during the evolution. The interplay between this directed propagation and mass-induced spin rotation gives rise to a two-dimensional trajectory exhibiting a rotational Zitterbewegung.

The experimental setup used in this work is shown in Fig.~\ref{fig1}(c).
The first two modes of the flute cavity, TE$_{101}$ and TE$_{102}$, serve as the two spatial dimensions of the simulated Dirac particle \cite{PhysRevLett.127.107701}. Their dispersive couplings to the transmon are $\frac{\chi_1}{2\pi} = 0.228\mathrm{MHz}$ and $\frac{\chi_2}{2\pi} = 0.063\mathrm{MHz}$. Thus, the two spatial dimensions are encoded in distinct modes of the same physical cavity, without requiring an additional cavity assembly. Additional experimental parameters are summarized in Table S1 of \cite{supp}.

\section{Experimental protocol}

\begin{figure*}[t]
    \centering
    \begin{subfigure}[t]{0.65\textwidth}
        \centering
        \includegraphics[width=\linewidth]{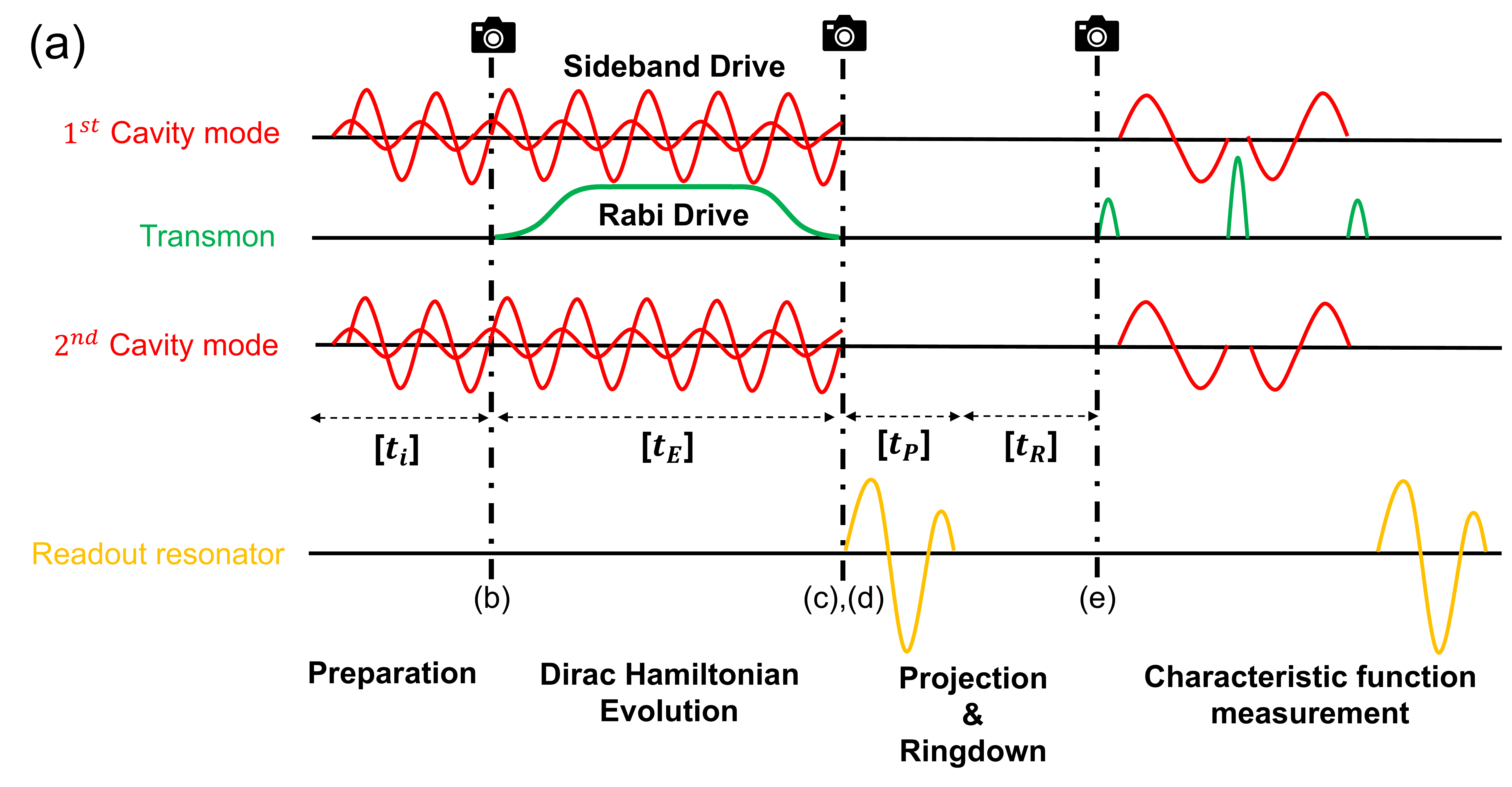}
    \end{subfigure}

    \begin{subfigure}[t]{0.95\textwidth}
        \centering
        \includegraphics[width=\linewidth]{Fig2_b_c.pdf}
    \end{subfigure}
    
    \begin{subfigure}[t]{0.95\textwidth}
        \centering
        \includegraphics[width=\linewidth]{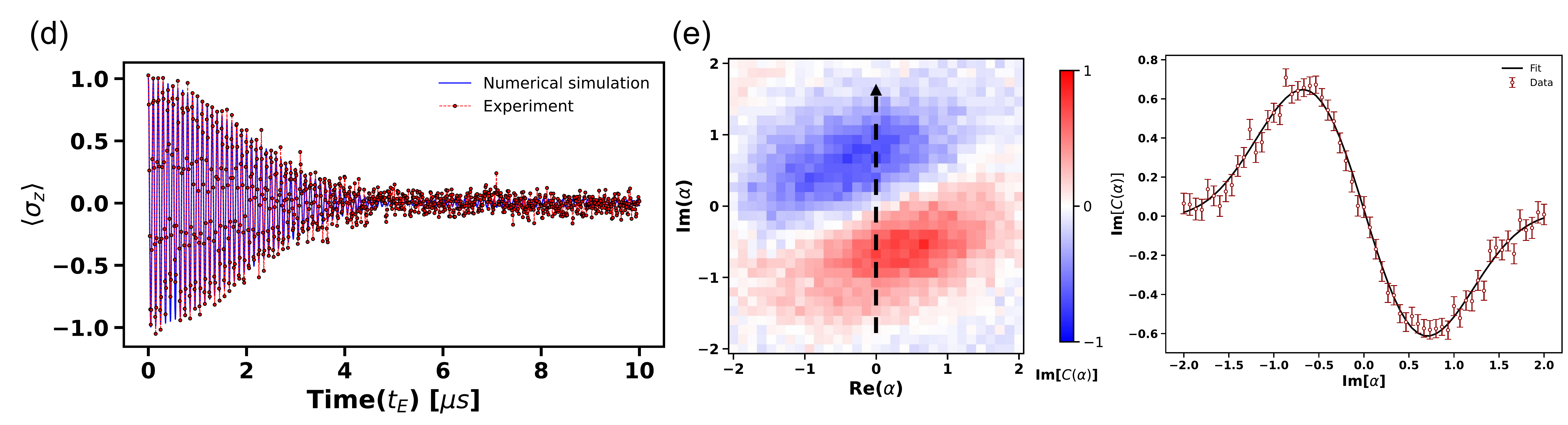}
    \end{subfigure}
    
    \caption{\justifying(a) Pulse sequence for the two-dimensional Zitterbewegung quantum simulations. The cavity state is initialized by double-sideband drives during the preparation stage (\(t_i\)), followed by Dirac Hamiltonian evolution under simultaneous Rabi and sideband drives (\(t_E\)). After qubit projection (\(t_P\)) and ringdown (\(t_R\)), the cavity state is measured via characteristic function measurements. The cavity-state snapshots corresponding to panels (b), (c), and (e) are marked in the sequence. (b) Cavity state snapshots during the preparation stage measured via Wigner tomography under a sideband drive with \(\Omega_{\mathrm{SB}} = 10~\mathrm{MHz}\). The cavity state oscillates along the imaginary axis of phase space with a period of \(0.1~\mu\mathrm{s}\). (c) Wigner tomography snapshots of the cavity states after \(2.075~\mu\mathrm{s}\) of evolution under Eq.~\eqref{1D_Dirac}, with the effective qubit projected onto the $\ket{+i}$ and $\ket{-i}$ states, respectively. The open symbols show the cavity state displacement of 0.5. The phase-space distributions are shown in the rotated frame compensating for cavity rotation during the projection and ringdown intervals. (d) Decaying dynamics of the qubit Rabi oscillation during the Dirac Hamiltonian evolution under Eq.~\eqref{1D_Dirac}. (e) Corresponding two-dimensional characteristic function and one-dimensional cut along the imaginary axis for the cavity state in the right panel of (c), measured after the projection and ringdown intervals.}
    \label{fig2}
\end{figure*}

The overall pulse sequence for the two-dimensional Zitterbewegung analog quantum simulations is illustrated in Fig.~\ref{fig2}(a). The experimental sequence consists of four stages: preparation, Dirac Hamiltonian evolution, projection and ringdown, and characteristic function measurement. In the following, we describe each stage based on single cavity mode dynamics and perform the necessary calibrations for the two-dimensional Zitterbewegung simulation.

Before applying the double-sideband drive in the preparation stage, the sideband tones must be properly calibrated. First, the double-sideband tones are balanced to ensure symmetric effective qubit--cavity interactions and suppress an undesired effective magnetic field effect \cite{wn7t-pyrq}. To compensate for frequency-dependent attenuation in the control lines, the qubit AC Stark shift induced by each sideband tone is independently measured, and the corresponding external drive powers are adjusted accordingly \cite{PhysRevA.74.042318, rdr}.

Figure~\ref{fig2}(b) shows the coherent-state dynamics of the cavity under a balanced double-sideband drive with $\Omega_{\mathrm{SB}}/2\pi=10~\mathrm{MHz}$, measured by Wigner tomography as a function of the $t_i$. The cavity displacement follows the relation $\frac{d\alpha_j(t)}{dt} = -i \epsilon_j(t)$. For a double-sideband drive of the form $\epsilon_j(t)=\alpha_j\Omega_{\mathrm{SB}}\cos(\Omega_{\mathrm{SB}}t)$, the resulting coherent state trajectory is therefore
\begin{equation}
\alpha_j(t)=-i\alpha_j\sin(\Omega_{\mathrm{SB}}t).
\label{sb_oscillation}
\end{equation}

Accordingly, the sideband amplitude $\alpha_j$ is calibrated from the maximum cavity displacement measured using the characteristic function at $t_i=1.025$ and $1.075~\mu\mathrm{s}$ in Fig.~\ref{fig2}(b), yielding $\alpha_1=0.67$ for the first cavity mode. An independent calibration using the sideband-induced AC Stark shift is given in Ref.~\cite{supp}. The sinusoidal trajectory along the imaginary axis, with no discernible displacement along the orthogonal quadrature, further confirms the balanced sideband drive. After these calibrations, the double-sideband drive is applied for a duration $t_i$ to prepare the initial cavity state, with $t_i$ setting the effective phase $\delta_j=\Omega_{\mathrm{SB}}t_i$ and thereby the initial condition for the Dirac Hamiltonian evolution stage \cite{wn7t-pyrq}.

Subsequently, a Rabi drive with a frequency $\Omega_{\mathrm{R}}$ is simultaneously applied with the double-sideband drives for an evolution time $t_E$. The qubit drive frequency is adjusted to compensate for the AC Stark shift induced by the double-sideband drives. When the Rabi frequency is resonant with the sideband frequency ($\Delta\Omega=0$) and the effective phase is set to $\delta_1=2\pi n$, where $n$ is an integer, the momentum operator couples to $\sigma_y$, yielding the effective Hamiltonian
\begin{equation}
H=-\frac{i\chi_1\alpha_1}{4}
\left(\hat{a}_1-\hat{a}_1^\dagger\right)\sigma_y.
\label{1D_Dirac}
\end{equation}

During the evolution stage, the cavity state undergoes oscillatory motion along the imaginary axis induced by the double-sideband drive, as shown in Fig.~\ref{fig2}(b), while the effective Hamiltonian in Eq.~\eqref{1D_Dirac} simultaneously generates a qubit-state-dependent displacement along the real axis, corresponding to the conditional displacement $D\left[(\chi_1\alpha_1/4)t_E\sigma_y\right]$. For the initial state $\ket{+,0}=\frac{1}{\sqrt{2}}\left(\ket{+i}+\ket{-i}\right)\ket{0}$, the two $\sigma_y$ eigenstates, $\ket{+i}$ and $\ket{-i}$, therefore undergo conditional displacements in opposite directions along the real axis in phase space. The resulting dynamics are visualized by the Wigner tomography shown in Fig.~\ref{fig2}(c) ($\frac{\chi_1\alpha_1}{4}t_E\sim0.5$). This conditional displacement progressively entangles the qubit and cavity, causing the qubit state purity to approach 0.5 with increasing evolution time, as shown in Fig.~\ref{fig2}(d). Since the decay of the Rabi oscillation is sensitive to the effective phase $\delta_j$, the measured Rabi dynamics also provide a means to calibrate $\delta_j$. The observed behavior shows good agreement with numerical simulations, supporting the calibrated simulation parameters. Additional Rabi dynamics under different conditions are shown in Ref.~\cite{supp}.

To disentangle the qubit--cavity system and obtain the cavity state conditioned on the qubit state, the entangled state is projected onto a selected qubit state. To minimize measurement backaction during this projection process, the DRACHMA readout protocol is employed, which suppresses the residual photon population in the readout resonator \cite{DRACHMA}. In addition, an extra $t_R$ is introduced after the measurement to ensure sufficient decay of any remaining photons.

After the projection, the position expectation value, $\langle \hat{X}_j \rangle=\frac{1}{2}\langle \hat{a}_j + \hat{a}_j^\dagger \rangle=\mathrm{Re}\big(\langle \hat{a}_j \rangle\big)$, is extracted from the qubit-state-dependent characteristic functions. The total characteristic function is reconstructed as $C_j(\alpha)=P_gC_{j,g}(\alpha)+P_eC_{j,e}(\alpha)$, where $P_{g/e}$ are the qubit projection probabilities and $C_{j,g/e}(\alpha)$ are the corresponding conditional cavity characteristic functions. The position expectation value is then obtained directly from the slope at the origin,
\begin{equation}
\mathrm{Re}(\langle \hat{a}_j \rangle) = \frac{1}{2}\left(\frac{\partial \mathrm{Im}[C_j(\alpha)]}{\partial \mathrm{Im}(\alpha)}\right)_{\alpha=0}
\end{equation}
avoiding the need for full two-dimensional characteristic-function measurements \cite{supp, OLIVARES2021127720}.

Figure~\ref{fig2}(e) illustrates the qubit-state-dependent characteristic functions after the projection and ringdown stage, together with representative one-dimensional cuts used to extract the slope. During $t_{\mathrm{rot}}=t_P+t_R$, the dispersive interaction induces an additional qubit-state-dependent phase-space rotation of the cavity state. The effect of this rotation is accounted for in post-processing when extracting $\langle\hat{X}_j\rangle$, such that the resulting expectation value corresponds to the cavity state immediately after the Dirac Hamiltonian evolution.

Applying these calibration procedures to both cavity modes establishes the experimental parameters required for the two-dimensional Zitterbewegung simulation. After preparing the desired initial qubit--cavity state, the position expectation value $\langle\hat{X}_j(t_E)\rangle$ is measured as a function of the evolution time $t_E$ for each cavity mode. This procedure reconstructs the two-dimensional trajectory of the simulated Dirac particle, revealing its Zitterbewegung dynamics.

\section{Results}
\subsection{1D Zitterbewegung}

\begin{figure}
    \centering
    \includegraphics[width = 0.48\textwidth]{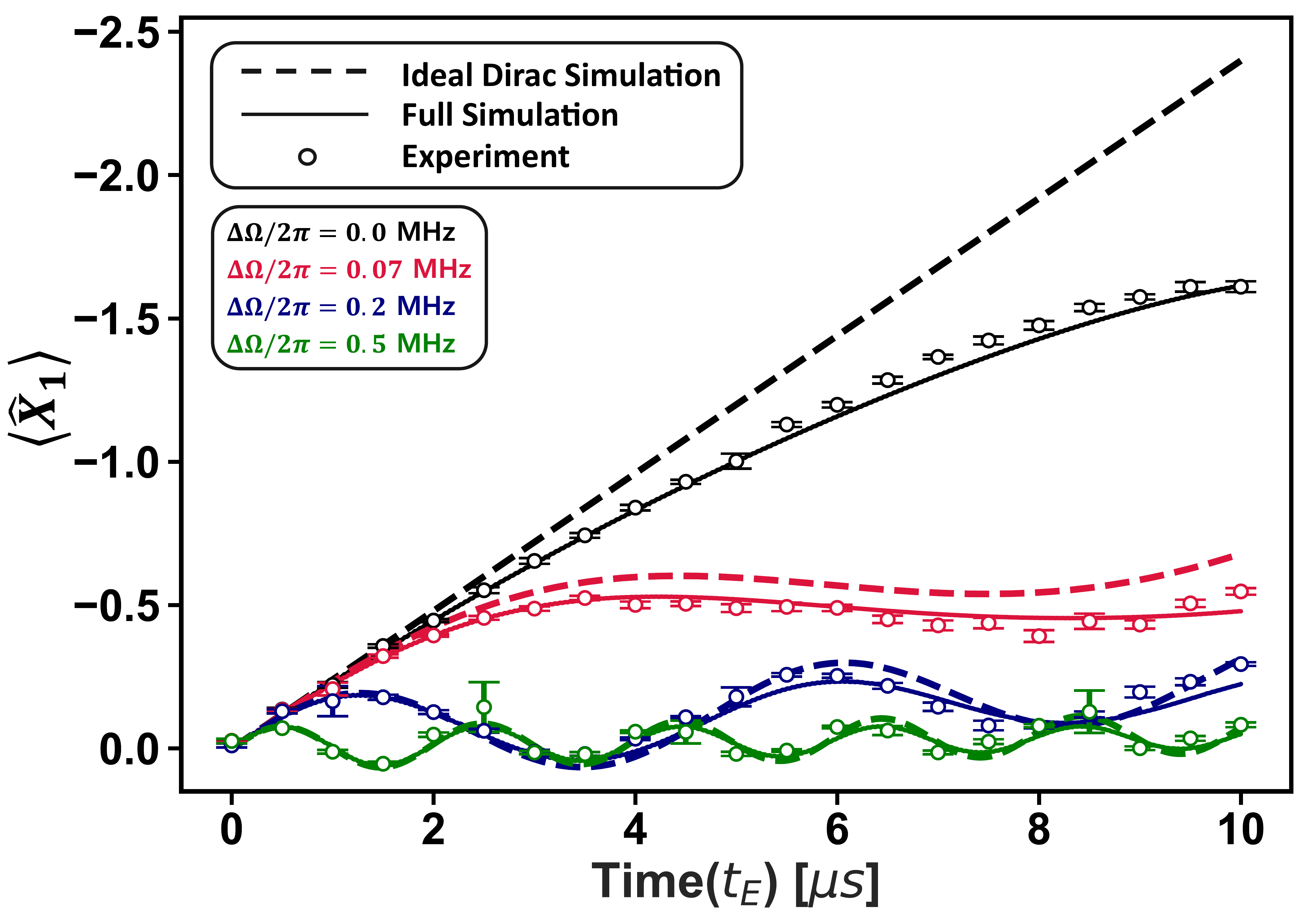}
    \caption{\justifying Experimental data and numerical simulations of one-dimensional Zitterbewegung dynamics for different effective particle masses. Dashed curves represent numerical simulations based on the ideal free Dirac Hamiltonian, while solid curves correspond to simulations based on the effective free Dirac Hamiltonian implemented in the circuit-QED system, including the effects of system decoherence, RWA breakdown, and the finite anharmonicity of the transmon qubit. Dotted markers denote the experimental data. The effective particle mass increases for the black, red, navy, and green curves, corresponding to $\Delta\Omega/2\pi = 0$, $0.07$, $0.2$, and $0.5~\mathrm{MHz}$, respectively.}
    \label{fig3}
\end{figure}

For the one-dimensional case, the free Dirac Hamiltonian is implemented using only the interaction between the effective qubit and the first cavity mode. The resulting effective Hamiltonian is given by
\begin{equation}
H = \frac{i \chi_1 \alpha_1}{4} \,(\hat{a}_1-\hat{a}_1^\dagger)\sigma_x+ \frac{\Delta\Omega}{2}\,\sigma_z
\label{1d_zitterbewegung}
\end{equation}
from Eq.~\eqref{Dirac_Hamiltonian}. Fig.~\ref{fig3} shows the quantum-simulated Zitterbewegung dynamics and numerical simulations of a one-dimensional free Dirac particle for several values of the effective mass parameter. For these simulations, the $\delta_1$ is adjusted to $\pi/2$ by controlling the preparation time of double sideband drive, and the initial state in the effective rotating frame is prepared as $\ket{+}\ket{\alpha=0}$. The sideband-drive frequency is fixed at $\Omega_{\mathrm{SB}}/2\pi = 10\,\mathrm{MHz}$ (so, $\Omega_R/2\pi = 10.0$, $10.07$, $10.2$, and $10.5\,\mathrm{MHz}$), and the double sideband drive amplitude is calibrated to $\alpha_1 = 0.67$. Accordingly, the effective speed of light is given by $c = -\frac{\chi_1 \alpha_1}{4} = -0.24\mu s^{-1},$
which corresponds to the slope of the black dashed curve in Fig.~\ref{fig3}.

The $\ket{+}$ and $\ket{-}$ spinor states generate cavity displacements in opposite directions, as illustrated in Fig.~\ref{fig1}(b). The effective mass term continuously rotates the spinor, periodically reversing the direction of the spin-dependent displacement and thereby producing Zitterbewegung. As the effective particle mass increases, the spinor rotates more rapidly, resulting in a higher Zitterbewegung frequency and a smaller oscillation amplitude as the system approaches the non-relativistic limit, consistent with the experimental results shown in Fig.~\ref{fig3}.

The deviations from the ideal Dirac dynamics observed in the full numerical simulations arise primarily from three effects. The dominant contribution is the finite spin-locking coherence time, $T_{2\rho}$, of the effective qubit under simultaneous continuous $10~\mathrm{MHz}$ Rabi driving and cavity double-sideband driving detuned by $\pm10~\mathrm{MHz}$ from the cavity resonance. The experimentally measured value of $T_{2\rho}$ was incorporated into the numerical master-equation simulations \cite{Lu2022}. Increasing $T_{2\rho}$ drives the dynamics toward the ideal Dirac limit, indicating spin-locking decoherence as a primary limitation of the effective implementation ~\cite{supp}. Two additional sources of deviation are the breakdown of the RWA and the finite anharmonicity of the transmon. In particular, the interaction term $H = \frac{\chi_1}{2} \hat{a}_1^\dagger \hat{a}_1 \left(\sigma^+ e^{i\Omega_{\mathrm{SB}} t_E} + \sigma^- e^{-i\Omega_{\mathrm{SB}} t_E}\right)$ is neglected under the RWA, but its influence becomes increasingly significant at longer evolution times and larger accumulated photon numbers \cite{wn7t-pyrq}. In addition, the finite anharmonicity of the transmon gives rise to multilevel effects beyond the two-level approximation, introducing further deviations from the ideal Dirac model.

Despite these limitations of the effective implementation, the experimental results are in good agreement with the full master-equation simulations. The remaining discrepancies between experiment and simulation are attributed mainly to temporal fluctuations of $T_{2\rho}$ (the experimentally observed $T_{1\rho}$ is sufficiently long that its fluctuations can be safely neglected), slow drifts of the Rabi frequency during the evolution, finite precision in the determination of the Rabi frequency, and calibration errors in the phase $\delta_j$, which depends sensitively on the chosen preparation time $t_i$. Nevertheless, for larger effective masses, corresponding to $\Delta\Omega/2\pi = 0.2$ and $0.5~\mathrm{MHz}$, the experimental data remain remarkably close to those predicted by the ideal free Dirac Hamiltonian, demonstrating the robustness of the proposed analog quantum simulation of relativistic Dirac dynamics.

\subsection{2D Zitterbewegung}

\begin{figure*}[t]
    \centering
    \includegraphics[width = 0.95\textwidth]{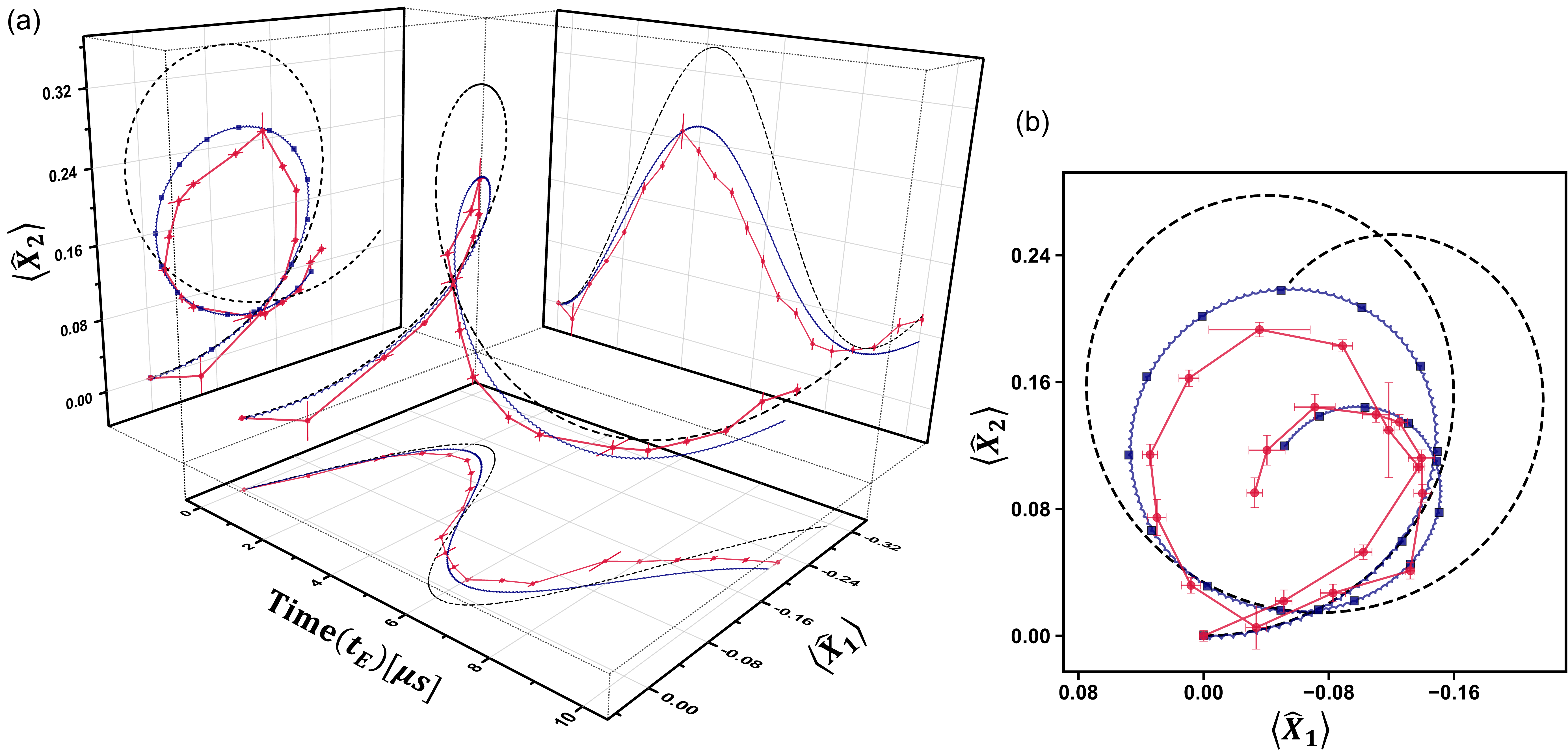}
    \caption{\justifying Experimental data and numerical simulations of two-dimensional Zitterbewegung for two effective particle masses. Black dashed lines show the ideal Dirac dynamics, while navy solid lines show simulations including experimental conditions, with navy squares indicating time points sampled every $0.5~\mu\mathrm{s}$. Red circles show the experimental data sampled at the same time intervals up to $10~\mu\mathrm{s}$. (a) Dynamics for $\Delta\Omega/2\pi=0.10~\mathrm{MHz}$, shown as the two orthogonal position components versus evolution time and their projection onto the two-dimensional position plane. (b) Two-dimensional trajectory for $\Delta\Omega/2\pi=0.15~\mathrm{MHz}$.}
    \label{fig4}
\end{figure*}

For the two-dimensional Zitterbewegung experiment, a single effective qubit is simultaneously coupled to two cavity modes via double sideband drives applied to each mode. The resulting effective Hamiltonian is given by,
\begin{equation}
H=\frac{i \chi_1 \alpha_1}{4} \, \sigma_{x} (\hat{a}_1 - \hat{a}_1^\dagger) 
-\frac{i \chi_2 \alpha_2}{4} \, \sigma_{y} (\hat{a}_2 - \hat{a}_2^\dagger)+
\frac{\Delta \Omega}{2} \, \sigma_z .
\end{equation}

To implement the two-dimensional free Dirac Hamiltonian of Eq.~\eqref{Dirac_Hamiltonian}, $\delta_1$ and $\delta_2$ are set to $\pi/2$ and $0$, respectively, by adjusting the preparation time of the double-sideband drives applied to each cavity mode. The initial state in the effective rotating frame is prepared as $\ket{+,0,0}$. The sideband-drive frequencies for both cavity modes are fixed at $\Omega_{\mathrm{SB},1}/2\pi=\Omega_{\mathrm{SB},2}/2\pi=10~\mathrm{MHz}$, while the Rabi frequencies in the Stark-shifted frame are chosen as $\Omega_{\mathrm{R}}/2\pi=10.10$ and $10.15~\mathrm{MHz}$. The double-sideband drive amplitudes are calibrated to $\alpha_1=0.44$ and $\alpha_2=1.6$ for the first and second cavity modes, respectively. Under these conditions, the effective speeds of light along the two simulated spatial dimensions become nearly identical, $c_{x,y}=\mp\frac{\chi_{1,2}\alpha_{1,2}}{4}=\mp0.157\mu s^{-1}$, thereby enabling the implementation of a two-dimensional free Dirac Hamiltonian.

Fig.~\ref{fig4} compares the quantum-simulation results of the Zitterbewegung dynamics of a two-dimensional free Dirac particle with the corresponding numerical simulations. Panel (a) shows the evolution of the two-dimensional trajectory $(\langle \hat{X}_1\rangle,\langle \hat{X}_2\rangle)$ for $\Delta\Omega/2\pi=0.10~\mathrm{MHz}$ as a function of the Dirac Hamiltonian evolution time $t_E$, together with its projections onto the two position--time planes and the two-dimensional position plane. Panel (b) shows the trajectory projected onto the two-dimensional position plane for $\Delta\Omega/2\pi=0.15~\mathrm{MHz}$.

The two-dimensional dynamics are a natural extension of the one-dimensional case. As illustrated in Fig.~\ref{fig1}(b), The effective mass term rotates the spinor and thereby continuously changes the velocity direction in the two-dimensional plane, producing the rotational Zitterbewegung trajectory. Increasing the effective particle mass leads to faster rotational motion with a smaller trajectory radius. As the evolution proceeds, the growing entanglement between the effective qubit and the cavity modes gradually reduces the visibility of the rotational motion. This behavior is consistent with the experimental observations in Fig.~\ref{fig4}(b).

Similar sources of error are also present in the two-dimensional simulations. However, because the effective qubit simultaneously couples to the momentum operators of two cavity modes along orthogonal spin axes, the resulting dynamics are expected to be more sensitive than in the one-dimensional case to imperfections such as finite spin-locking coherence times ($T_{1\rho}$ and $T_{2\rho}$) under double-sideband driving and fluctuations in the Rabi-drive amplitude. Nevertheless, the qualitative agreement between the experimental results and the numerical simulations in Fig.~\ref{fig4} suggests that further improvements in the intrinsic coherence of the effective qubit, together with enhanced stability of the control electronics (e.g., reduced voltage fluctuations), should enable an even closer realization of the ideal two-dimensional Zitterbewegung dynamics in the proposed analog quantum simulation.

\section{Conclusion}

In conclusion, we have experimentally realized two-dimensional free massive Dirac dynamics and observed rotational Zitterbewegung in a programmable multimode circuit-QED simulator. Using a single Rabi-driven effective qubit coupled to two bosonic cavity modes, we observe the dependence of the rotational trajectory on the effective Dirac mass. Importantly, the two-dimensional dynamics are not simply a combination of two independent one-dimensional motions. Because the two bosonic modes, representing orthogonal momentum degrees of freedom, couple to non-commuting spin components of the same effective qubit, the shared qubit degree of freedom gives rise to correlated multimode dynamics that cannot be described as a simple superposition of independent cavity displacements. These coupled spin–momentum dynamics are directly reflected in the measured two-dimensional Zitterbewegung trajectories.

The observed trajectories are well reproduced by master-equation simulations incorporating measured decoherence and device nonidealities. The primary limitation is the finite spin-locking coherence of the Rabi-driven qubit, while additional deviations arise from corrections beyond the RWA, finite transmon anharmonicity, and control-parameter fluctuations. This close agreement between experiment and simulation suggests that mitigating these limitations should enable higher-fidelity quantum simulations approaching the ideal Dirac dynamics.

As the first experimental realization of free two-dimensional massive Dirac dynamics, this work establishes a versatile platform for analog quantum simulations of relativistic and Dirac-like physics. Its programmability opens the way to simulations in external electromagnetic fields, including magnetic-field-induced dynamics and Klein tunneling, as well as extensions toward topological Dirac physics \cite{Novoselov2005, RevModPhys.82.1959}. More broadly, our results highlight multimode circuit-QED systems as resource-efficient and extensible platforms for exploring complex quantum dynamics.

\begin{acknowledgments}
This research was supported in part by Creation of the Quantum Information Science R\&D Ecosystem (Based on Human Resources) [RS-2023-00256050], the Quantum Technology R\&D Leading Program (Quantum Computing) [RS-2023-00282500], and Quantum Science and Technology Flagship Project (Quantum Computing) [RS-2025-25464760] through the National Research Foundation of Korea (NRF), funded by the Korean government (Ministry of Science and ICT (MSIT)). This research was supported by the Israeli Science Foundation grant No. 657/23, and Technion’s Helen Diller Quantum Center.
\end{acknowledgments}

\bibliographystyle{apsrev4-2}
\bibliography{refs}

\clearpage

\setcounter{section}{0}
\renewcommand{\thesection}{S\arabic{section}}

\setcounter{figure}{0}
\renewcommand{\thefigure}{S\arabic{figure}}

\setcounter{table}{0}
\renewcommand{\thetable}{S\arabic{table}}

\setcounter{equation}{0}
\renewcommand{\theequation}{S\arabic{equation}}

\clearpage
\begin{center}
{\large\bfseries Supplemental Material}
\end{center}
\FloatBarrier

\section{Methods and calibrations} \label{Appendix A}

\subsection{Drive frequency dependent attenuation calibration}

\begin{figure}
    \includegraphics[width = 0.45\textwidth]{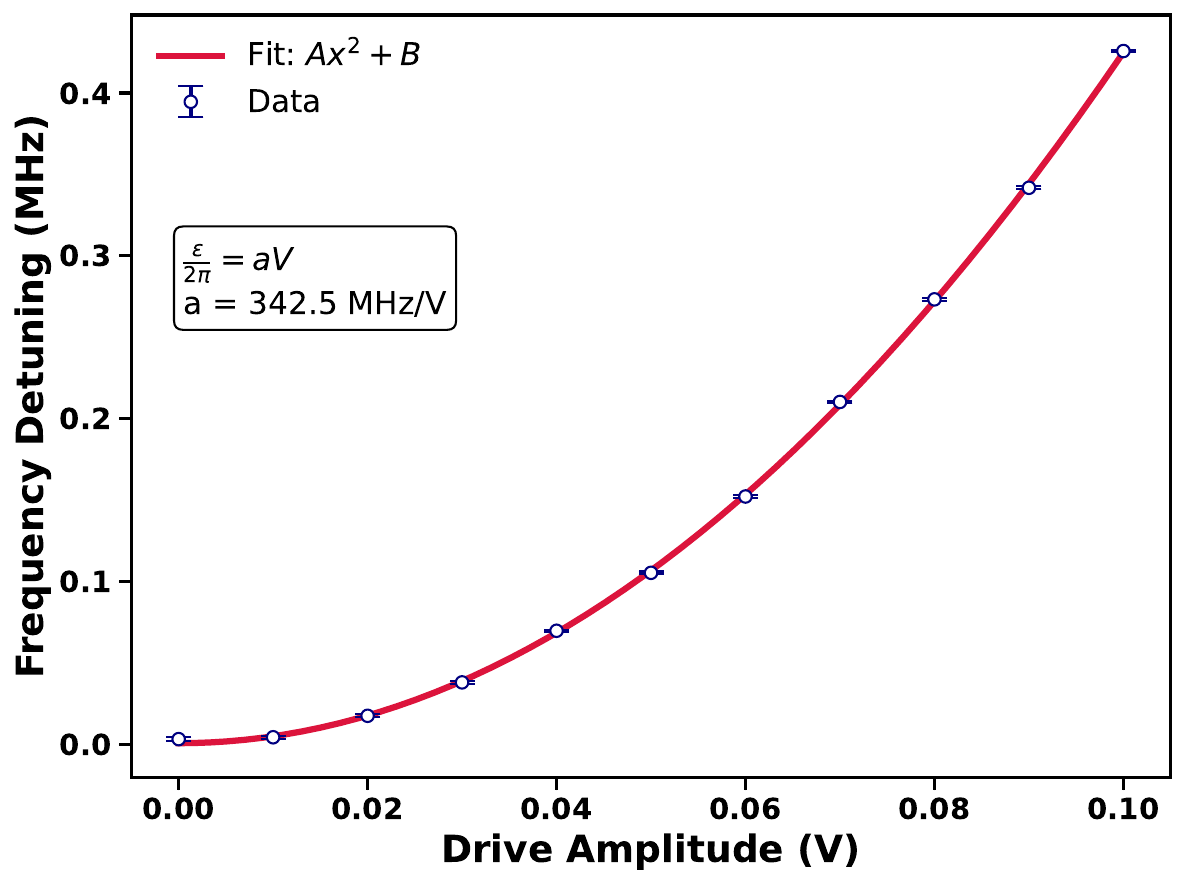}
    \caption{\justifying AC Stark shift versus cavity-mode SSB drive voltage and extraction of the attenuation coefficient.}
    \label{fig_ramsey_amp_sweep}
\end{figure}

To realize symmetric double-sideband drives with equal amplitudes at frequencies detuned by \(\pm\Omega_{\mathrm{SB}}\) from the cavity resonance, drive frequency dependent attenuation calibration is required. Although the two sideband tones are delivered through the same RF line, frequency-dependent attenuation in the control electronics and wiring can result in unequal effective drive amplitudes at the cavity input. To compensate for this effect, the attenuation coefficient \(a\) in the relation $\frac{\epsilon}{2\pi}=aV$ is calibrated for each sideband frequency. Here, \(\epsilon/2\pi\) denotes the effective drive amplitude experienced by the cavity mode, while \(V\) is the drive voltage generated by the experimental electronics.

The calibration is performed by measuring the AC Stark shift of the qubit as a function of the applied single-sideband drive voltage. The resulting Stark shift is given by

\begin{equation}
\Delta f = \frac{1}{2\pi}\frac{2\chi}{\frac{\kappa^2}{4}+\Omega_{\mathrm{SB}}^2}\left(\frac{\frac{\kappa^2}{4}-\chi^2+\Omega_{\mathrm{SB}}^2}{\frac{\kappa^2}{4}+\chi^2+\Omega_{\mathrm{SB}}^2}\right)a^2V^2 .
\end{equation}

This expression is derived in Ref.~\cite{PhysRevA.74.042318}. The attenuation coefficients corresponding to the two sideband frequencies, detuned by \(+\Omega_{\mathrm{SB}}\) and \(-\Omega_{\mathrm{SB}}\) from the cavity resonance, are independently extracted and subsequently used to compensate the applied drive amplitudes in the experiment.

\subsection{Sideband amplitude calibration}

\begin{figure}
    \centering
    \begin{subfigure}{\linewidth}
        \centering
        \includegraphics[width=0.85\textwidth]{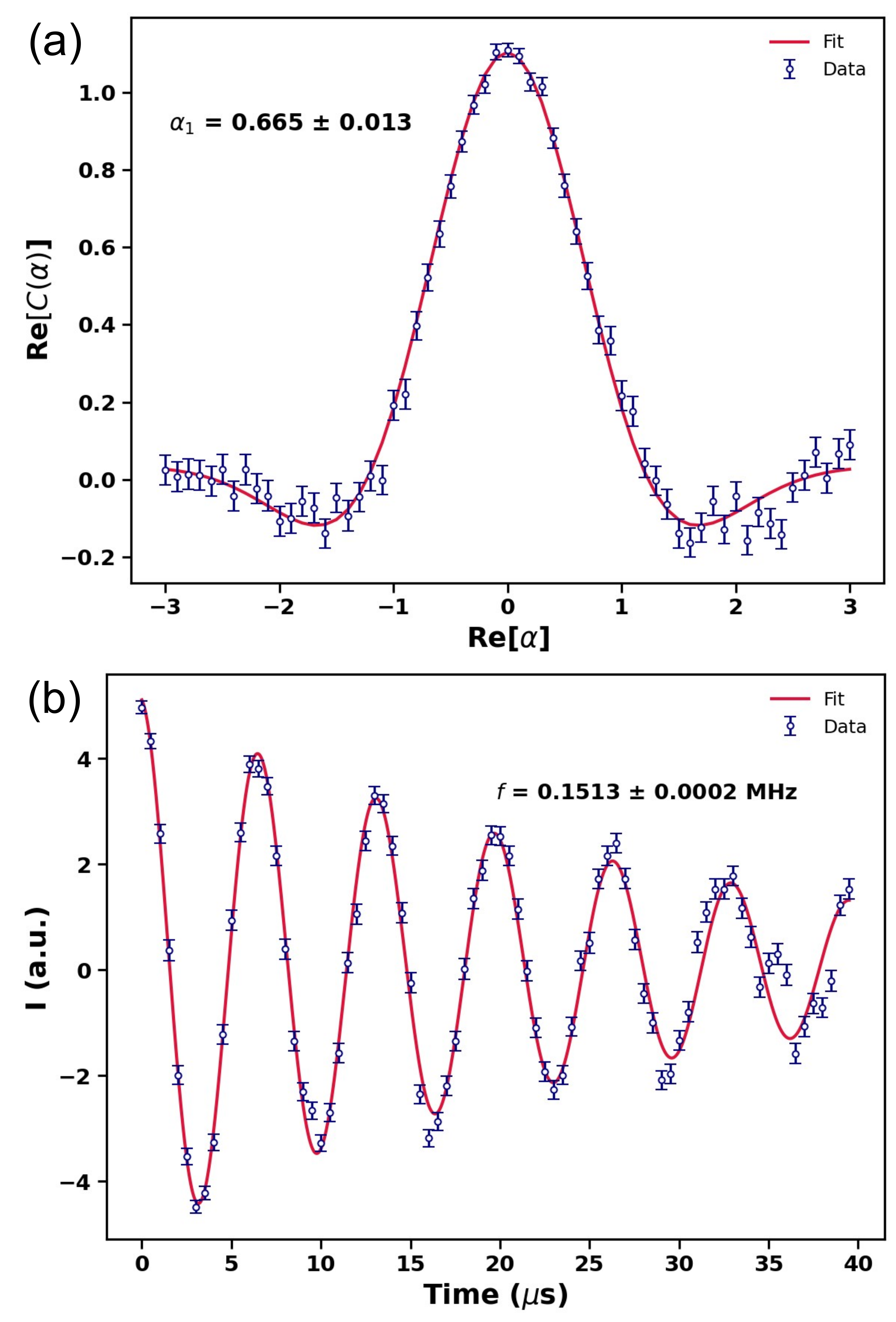}
        \label{fig:characteristic-function_ramsey_SB}
    \end{subfigure}
    \caption{\justifying (a) Calibration of the cavity-mode double-sideband drive amplitude using the real part of the single-line characteristic function measured at the maximum coherent-state displacement. (b) Ramsey measurement of the qubit ac Stark shift under the same symmetric double-sideband driving conditions as in (a).}
    \label{fig:sb-calibration}
\end{figure}

As shown in Eq.~(3) of the main text, balancing the effective amplitudes of the double-sideband drives leads to coherent-state oscillations along a single direction in the IQ plane.

The cavity-mode double-sideband drive strength is calibrated by measuring the single-line characteristic function at the time of maximum coherent state displacement and fitting the result to the theoretical expression below \cite{10.1093/acprof:oso/9780198509141.005.0001}. 
\begin{equation}
C^{\ket{\alpha_1}}(\alpha)=\bra{\alpha_1}\hat{D}(\alpha)\ket{\alpha_1}=C^{\ket{0}}(\alpha)e^{(2iIm(\alpha\alpha_1^*))}
\end{equation}
where $C^{\ket{0}}(\alpha)=\bra{0}\hat{D}(\alpha)\ket{0}=e^{\big(-\frac{|\alpha|^2}{2}\big)}$. (We use a characteristic function in symmetric order.) For a 10 MHz oscillation, the maximum displacement occurs at ($t = 0.025, 0.075\mu\mathrm{s}$). Fig.~\ref{fig:sb-calibration}(a) presents the real part of the characteristic function obtained from a snapshot of the measurement sequence at $t = 1.025\mu\mathrm{s}$ in Fig.~2(b) of the main text.

As an independent calibration, the symmetric sideband amplitude $\alpha_1$ is extracted from the ac Stark shift induced by the cavity double-sideband drive as shown in Fig.~\ref{fig:sb-calibration}(b). Following Ref.~\cite{wn7t-pyrq, Hacohen2016}, $\Delta f = \frac{\chi_1}{2}\alpha_1^2$. From the Ramsey measurement, we obtain $\Delta f = 0.0513~\mathrm{MHz}$, where a $0.1~\mathrm{MHz}$ detuning is intentionally introduced to improve the frequency fit, yielding $\alpha_1 = 0.67$. This value agrees well with that obtained from the characteristic function measurement in Fig.~\ref{fig:sb-calibration}(a), confirming the sideband amplitude calibration.

\subsection{Momentum - spinor coupling axis calibration}

\begin{figure}
    \includegraphics[width=0.45\textwidth]{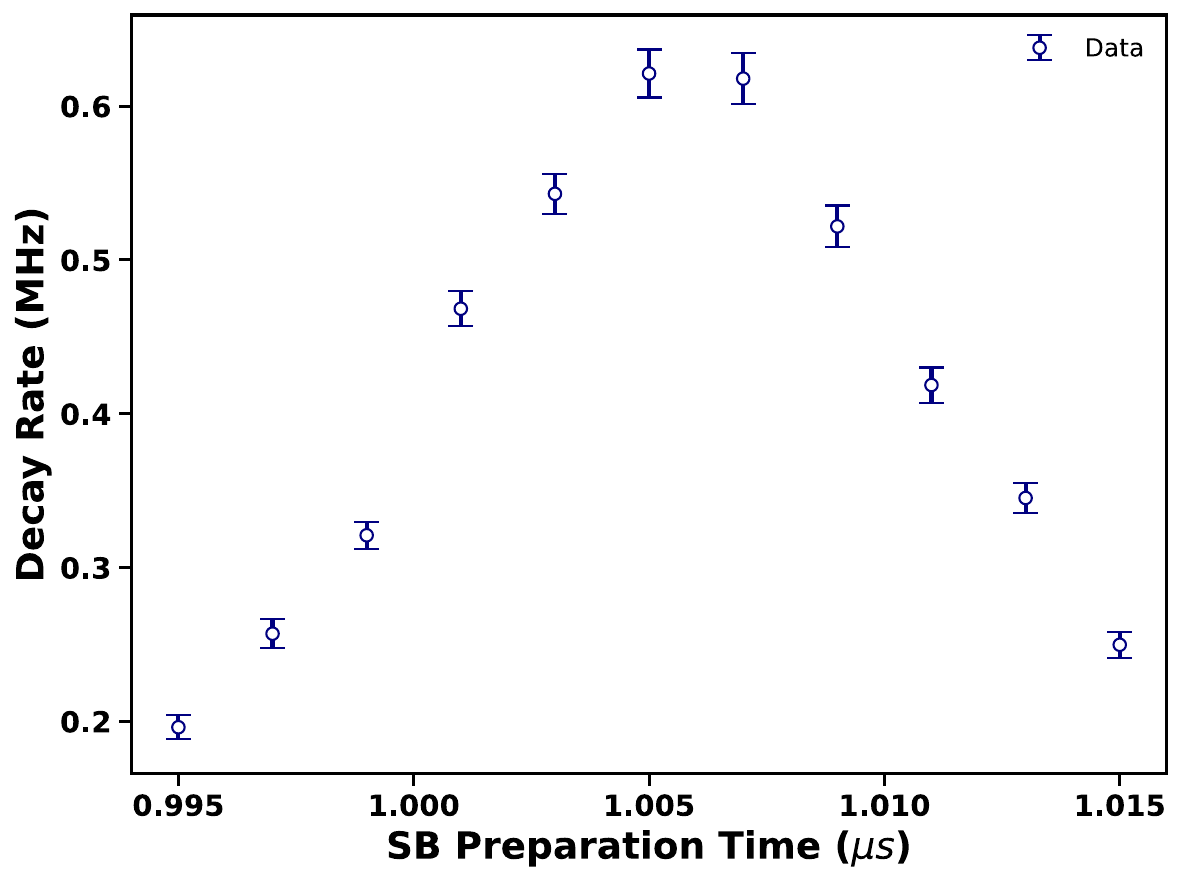}
    \caption{\justifying Calibration of $\delta_j$ using the decay rate of qubit Rabi oscillations measured as a function of the double sideband preparation time in the preparation stage. The maximum decay rate occurs at $t_i = 1.005~\mu\mathrm{s}$, where the phase \(\delta_j=0\).}
    \label{Rabi_decay_rate_vs_steady_time_sweep}
\end{figure}

As described in the main text, during the preparation stage the double sideband drive is applied for a preparation time ($t_i$), allowing the cavity field to reach the desired initial state for the subsequent Dirac Hamiltonian evolution. This preparation process effectively determines the phase parameter \(\delta_j = \Omega_{SB}t_i\).

However, in the actual experiment, an additional timing consideration is required. Specifically, the arrival times of the cavity double sideband drive and the qubit Rabi drive must be synchronized while accounting for the finite ramp-up time of the Rabi pulse. Therefore, \(\delta_j\) should be calibrated by taking these timing effects into account.

To characterize this synchronization condition, we measure the decay rate of the qubit Rabi oscillation while sweeping the double sideband preparation time. This allows us to determine along which axis the momentum operator couples to the effective qubit spinor. When the sideband and Rabi drives are resonant, i.e., \(\Delta\Omega = 0\), the effective Hamiltonian, Eq.~(2) of the main text, becomes
\begin{equation}
H = -\frac{i\chi_j\alpha_j}{4}(\hat{a}_j-\hat{a}_j^\dagger)\sigma_{\delta_j+\frac{\pi}{2}}.
\label{B3}
\end{equation}
Sweeping the sideband preparation time effectively sweeps the angle \(\delta_j\), thereby rotating the momentum--spinor coupling axis on the effective qubit Bloch sphere.

Assuming that the effective qubit is initially prepared in the \(|+\rangle\) state, i.e., the eigenstate of \(\sigma_x\), and the momentum--spinor coupling axis is aligned with the \(x\)-axis ($\delta_j=\pm\frac{\pi}{2}$), the effective Hamiltonian Eq.~\eqref{B3} acts as an unconditional displacement operator. In contrast, when the coupling axis is aligned with the \(y\)-axis ($\delta_j=0,\pi$), the Hamiltonian acts as a conditional displacement operator, resulting in progressive entanglement between the effective qubit and the cavity state during the evolution. Such qubit--cavity entanglement reduces the purity of the effective qubit state, which experimentally appears as an enhanced decay of the Rabi oscillation. Accordingly, the closer the momentum--spinor coupling axis is to the \(y\)-axis, the faster the entanglement develops and the larger the observed Rabi oscillation decay rate becomes as shown in Fig.~\ref{Rabi_decay_rate_vs_steady_time_sweep}. The sideband preparation time at which the decay rate reaches its maximum therefore identifies the synchronization condition corresponding to a momentum--spinor coupling axis aligned with the \(y\)-direction.

\subsection{Extraction of $\langle \hat{X}_1 \rangle$ and $\langle \hat{X}_2 \rangle$}

\subsubsection{Characteristic function - $\langle \hat{X} \rangle$ measurement}

In 1D experiment, one-dimensional characteristic-function measurements along the \(\mathrm{Re}(\alpha)\) and \(\mathrm{Im}(\alpha)\) axes are used to extract the expectation value of the position operator, \(\langle \hat{X}_1 \rangle = \frac{1}{2}\langle \hat{a}_1 + \hat{a}_1^\dagger \rangle = \mathrm{Re}[\langle \hat{a}_1 \rangle]\). The characteristic function is defined as
\begin{equation}
C(\alpha) \equiv \mathrm{Tr}[\rho \mathcal{D}(\alpha)] = 1 + \alpha \langle \hat{a}_1^\dagger \rangle - \alpha^* \langle \hat{a}_1 \rangle + O(\alpha^2),
\label{B4}
\end{equation}
where $\mathcal{D}(\alpha) = e^{(\alpha \hat{a}_1^\dagger - \alpha^* \hat{a}_1)} = 1 + \alpha \hat{a}_1^\dagger - \alpha^* \hat{a}_1 + O(\alpha^2)$ \cite{10.1093/acprof:oso/9780198509141.005.0001}.
\begin{equation}
\begin{aligned}
\langle \hat{a}_1 \rangle
&= -\left.\frac{\partial C(\alpha)}{\partial \alpha^*}\right|_{\alpha=0} \\
&= -\frac{1}{2}\left.\left(\frac{\partial C(\alpha)}{\partial x}
+ i\frac{\partial C(\alpha)}{\partial y}\right)\right|_{\alpha=0}.
\end{aligned}
\end{equation}
where $C(\alpha)=\mathrm{Re}[C(\alpha)]+i\,\mathrm{Im}[C(\alpha)]$, $x = \mathrm{Re}(\alpha)$, $y = \mathrm{Im}(\alpha)$
\begin{equation}
\begin{aligned}
\mathrm{Re}(\langle \hat{a}_1 \rangle)
&=
-\frac{1}{2}\left.\left(
\frac{\partial \mathrm{Re}[C(\alpha)]}{\partial x}
-
\frac{\partial \mathrm{Im}[C(\alpha)]}{\partial y}
\right)\right|_{\alpha=0} \\
&=
\frac{1}{2}\left.
\frac{\partial \mathrm{Im}[C(\alpha)]}{\partial y}
\right|_{\alpha=0}.
\end{aligned}
\end{equation}
Since \(\mathrm{Re}[C(\alpha)]\) is an even function and therefore symmetric about the origin, \(\frac{\partial \mathrm{Re}[C(\alpha)]}{\partial x}\big|_{\alpha=0}=0\). Furthermore, because the density matrix in Eq.~\eqref{B4} is given by \(\rho = P_g \ket{c_g,g}\bra{c_g,g} + P_e \ket{c_e,e}\bra{c_e,e}\),
\begin{equation}
\begin{aligned}
C(\alpha)
&=
P_g \langle c_g,g|\mathcal{D}(\alpha)|c_g,g\rangle
+
P_e \langle c_e,e|\mathcal{D}(\alpha)|c_e,e\rangle \\
&=
P_g C_g(\alpha) + P_e C_e(\alpha)
\end{aligned}
\end{equation}
then, $\mathrm{Re}(\langle \hat{a}_1 \rangle) = \frac{P_g}{2}\frac{\partial \mathrm{Im}[C_g(\alpha)]}{\partial y}\big|_{\alpha=0} + \frac{P_e}{2}\frac{\partial \mathrm{Im}[C_e(\alpha)]}{\partial y}\big|_{\alpha=0}$.

\subsubsection{Characteristic function - $\langle \hat{X}_1 \rangle$ and $\langle \hat{X}_2 \rangle$ measurement}

As in the one-dimensional case, the expectation values of the position operators for the 2D are obtained as $\langle \hat{X}_1 \rangle=\frac{1}{2}\langle \hat{a}_1+\hat{a}_1^\dagger\rangle=\mathrm{Re}\!\left(\langle\hat{a}_1\rangle\right), \langle \hat{X}_2 \rangle=\frac{1}{2}\langle\hat{a}_2+\hat{a}_2^\dagger\rangle=\mathrm{Re}\!\left(\langle\hat{a}_2\rangle\right).$

The two-mode characteristic function is defined as
\begin{equation}
\begin{aligned}
C_J(\alpha,\beta)&=\mathrm{Tr}\!\left[\rho D_A(\alpha)D_B(\beta)\right] \\
&=1+\alpha\langle\hat{a}_1^\dagger\rangle-\alpha^*\langle\hat{a}_1\rangle+\beta\langle\hat{a}_2^\dagger\rangle-\beta^*\langle\hat{a}_2\rangle+\cdots,
\end{aligned}
\end{equation}
where $\alpha=x_A+i y_A$ and $\beta=x_B+i y_B$ \cite{Y_Lai_1989}. Therefore,
$\langle\hat{a}_1\rangle=-\left.\frac{\partial C_J}{\partial\alpha^*}\right|_{\alpha=\beta=0},
\langle\hat{a}_2\rangle=-\left.\frac{\partial C_J}{\partial\beta^*}\right|_{\alpha=\beta=0},$ which gives
\begin{equation}
\mathrm{Re}\!\left(\langle\hat{a}_{1/2}\rangle\right)=\frac{1}{2}\left.\frac{\partial\,\mathrm{Im}(C_J)}{\partial y_{A/B}}\right|_{\alpha=\beta=0}.
\end{equation}

\subsubsection{Compensation of phase space rotation}

Experimentally measured characteristic functions correspond to the cavity state after the projection and ringdown time, $t_{\mathrm{rot}} = t_P + t_R$. Therefore, in order to obtain the cavity observable \(\mathrm{Re}(\langle \hat{a_j} \rangle)\) immediately after the Dirac Hamiltonian evolution, the phase-space rotation induced by dispersive interaction during the $t_{\mathrm{rot}}$ must be compensated.

The relevant dispersive interaction term is $H_\chi = \frac{\chi_j}{2} \hat{a}_j^\dagger \hat{a}_j \sigma_z$ which causes the cavity coherent state to rotate in opposite directions depending on the qubit state. Under this Hamiltonian, the cavity state evolves as $
\alpha(t)=\alpha(0)e^{-i\theta},$ where $\theta=\frac{\chi_j t_{\mathrm{rot}}}{2}$. Since \(t_{\mathrm{rot}}\) is much shorter than the cavity photon lifetime in the present experiment, dissipative effect such as cavity decay is neglected, and only the phase-space rotation is considered.

Accordingly, the measured characteristic function $C(\alpha,t_0+t_{\mathrm{rot}})$ corresponds to a rotated version of $C(\alpha e^{+i\theta},t_0)$, where $t_0$ denotes the target measurement time at which the characteristic function is evaluated, corresponding to the end of the Dirac Hamiltonian evolution. Since the experimentally accessible quantity is the characteristic function at \(t_0+t_{\mathrm{rot}}\), an inverse rotation must be applied in order to recover the characteristic function at time \(t_0\). 
Defining $\alpha=x+iy$ and $\alpha'=\alpha e^{+i\theta}=x'+iy',$
the rotated coordinates become $x'=x\cos\theta-y\sin\theta, y'=x\sin\theta+y\cos\theta.$

Using the chain rule, the derivatives of the characteristic function at time \(t_0\) can be expressed in terms of the derivatives of the measured characteristic function at \(t_0+t_{\mathrm{rot}}\) as

\begin{equation}
\begin{aligned}
&\begin{pmatrix}
\partial_x\,\mathrm{Im}C(\alpha',t_0)\\
\partial_y\,\mathrm{Im}C(\alpha',t_0)
\end{pmatrix}
\\
&=
\begin{pmatrix}
\cos\theta & \sin\theta\\
-\sin\theta & \cos\theta
\end{pmatrix}
\begin{pmatrix}
\partial_x\,\mathrm{Im}C(\alpha,t_0+t_{\mathrm{rot}})\\
\partial_y\,\mathrm{Im}C(\alpha,t_0+t_{\mathrm{rot}})
\end{pmatrix}.
\end{aligned}
\end{equation}

All derivatives are evaluated at \(\alpha=0\). Since the origin is invariant under phase-space rotation, \(\alpha=0\) implies \(\alpha'=0\). Therefore, the real part of the annihilation operator expectation value is obtained from the characteristic function as $
\mathrm{Re}(\langle \hat{a}_j\rangle_{t_0})=\frac{1}{2}\left[
\frac{\partial \mathrm{Im}\,C(\alpha',t_0)}{\partial y}
\right]_{\alpha=0},$ which leads to

\begin{equation}
\begin{aligned}
\mathrm{Re}(\langle \hat{a}_j\rangle_{t_0})
=
\frac{1}{2}
\Big[
&-\sin\theta \frac{\partial \mathrm{Im}\,C(\alpha,t_0+t_{\mathrm{rot}})}{\partial x} \\
&+\cos\theta \frac{\partial \mathrm{Im}\,C(\alpha,t_0+t_{\mathrm{rot}})}{\partial y}
\Big]_{\alpha=0}
\end{aligned}
\end{equation}

Thus, by applying the inverse phase-space rotation to the gradient of the measured characteristic function, the dispersive rotation accumulated during the \textit{Projection} and \textit{Ringdown} time can be considered, allowing the reconstruction of \(\mathrm{Re}(\langle \hat{a}_j\rangle)\) at the time \(t_0\).

\section{Anharmonicity limitation} \label{Appendix B}

\begin{figure}
    \centering
    \includegraphics[width = 0.48\textwidth]{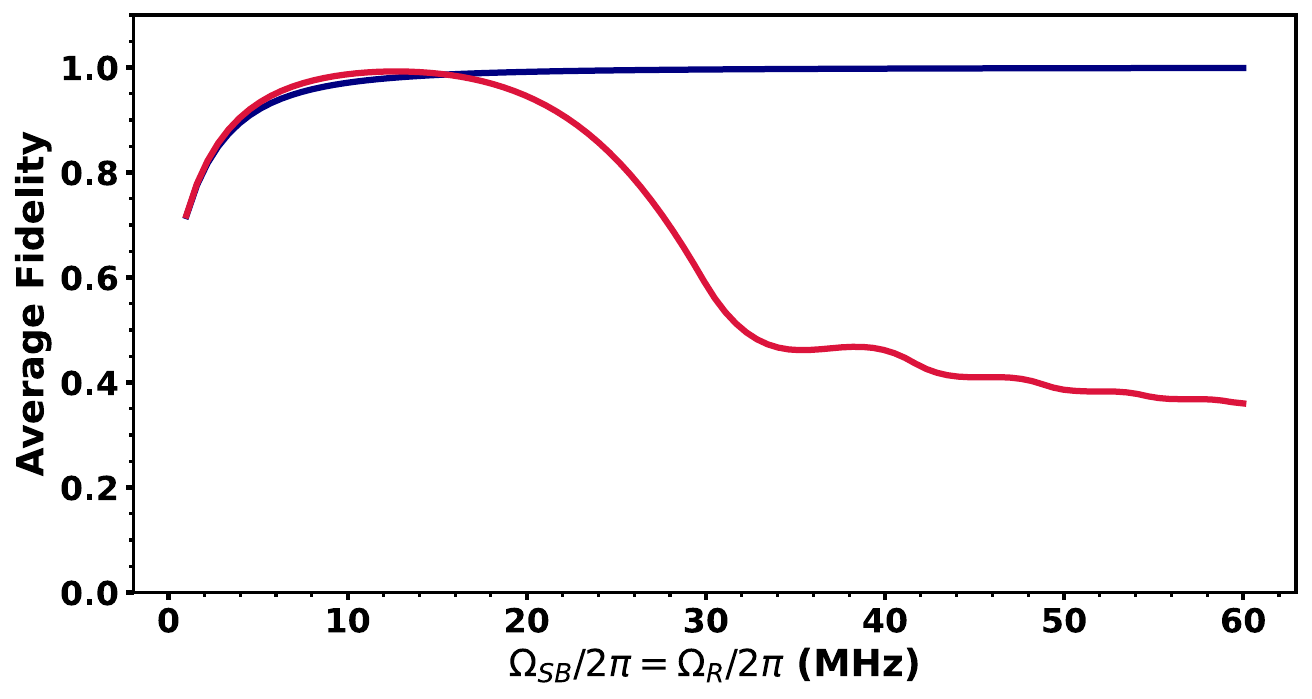}
    \caption{\justifying Time-averaged state fidelity with respect to the ideal Dirac Hamiltonian as a function of $\Omega_{\mathrm{SB}}$. The navy (red) curve corresponds to an ideal two-level qubit (a multilevel transmon with the experimentally relevant finite anharmonicity).}
    \label{fig_fidelity_anharm}
\end{figure}

Large sideband frequencies, $\Omega_{\mathrm{SB}}$, suppress the error arising from the rotating-wave approximation (RWA) \cite{wn7t-pyrq}. For a given effective Dirac mass, however, the Rabi frequency $\Omega_{\mathrm{R}}$ must remain close to $\Omega_{\mathrm{SB}}$. Consequently, increasing $\Omega_{\mathrm{SB}}$ requires a larger Rabi frequency, whose maximum value is ultimately limited by the finite anharmonicity of the transmon. 

To quantify this limitation, we simulate the qubit dynamics over a $10\mu\mathrm{s}$ evolution using the experimental parameters employed in the main text. Fig.~\ref{fig_fidelity_anharm} compares the time-averaged fidelity with the ideal Dirac Hamiltonian as a function of $\Omega_{\mathrm{SB}}$. The navy curve is obtained using the effective Dirac Hamiltonian derived for an ideal two-level qubit, whereas the red curve includes the experimentally relevant multilevel transmon with finite anharmonicity. 

As shown by the navy curve, the fidelity increases monotonically with $\Omega_{\mathrm{SB}}$, reflecting the suppression of the RWA error. In contrast, when the finite anharmonicity of the transmon is taken into account (red curve), the fidelity no longer improves monotonically. Although increasing $\Omega_{\mathrm{SB}}$ reduces the RWA error, it also requires a larger $\Omega_{\mathrm{R}}$, leading to enhanced leakage to higher transmon levels \cite{Babu2021}. Consequently, an optimal range of $\Omega_{\mathrm{SB}}$ exists where the RWA error and higher-level leakage are balanced. The experimentally adopted value, $\Omega_{\mathrm{SB}}/2\pi = 10\mathrm{MHz}$, lies close to this optimum. 

These results demonstrate that the finite anharmonicity of the transmon, $E_C/\hbar = 230\mathrm{MHz}$, fundamentally limits the fidelity with which the circuit-QED system can emulate the ideal Dirac Hamiltonian.

\section{Various rabi dynamics} \label{Appendix C}

\begin{figure}
    \centering
    \begin{subfigure}[t]{0.48\textwidth}
        \centering
        \includegraphics[width=\textwidth]{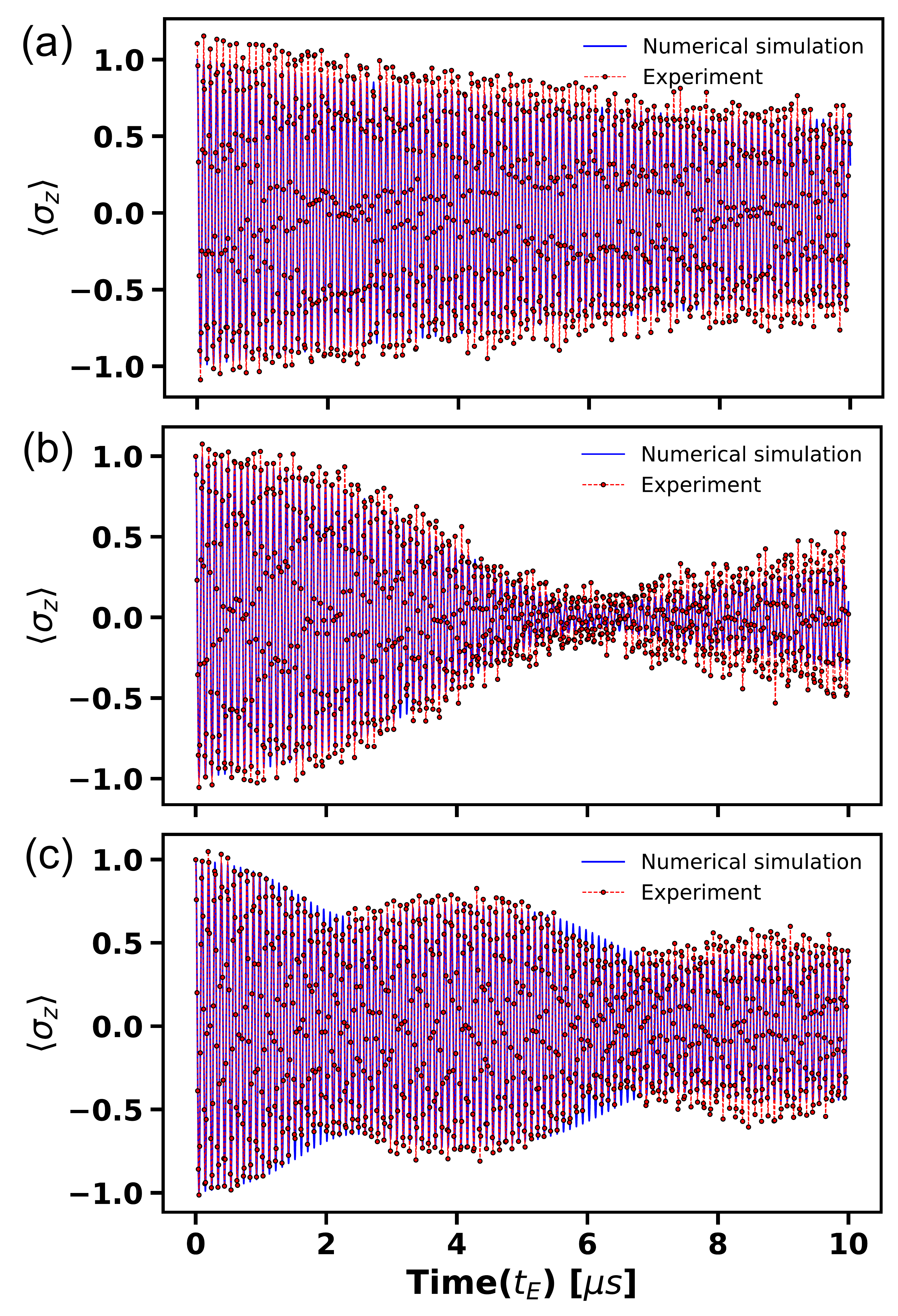}
    \end{subfigure}
    
    \caption{\justifying Effective qubit Rabi dynamics of the effective quantum Rabi Hamiltonian in the ultrastrong-coupling regime. From (a) to (c), the panels correspond to effective qubit transition frequencies of $\Delta\Omega/2\pi = 0$, $0.07$, and $0.2~\mathrm{MHz}$ in Eq.~(6) of the main text.}
    \label{fig_various_rabi_dynamics}
\end{figure}

\begin{figure}
    \centering
    \includegraphics[width = 0.45\textwidth]{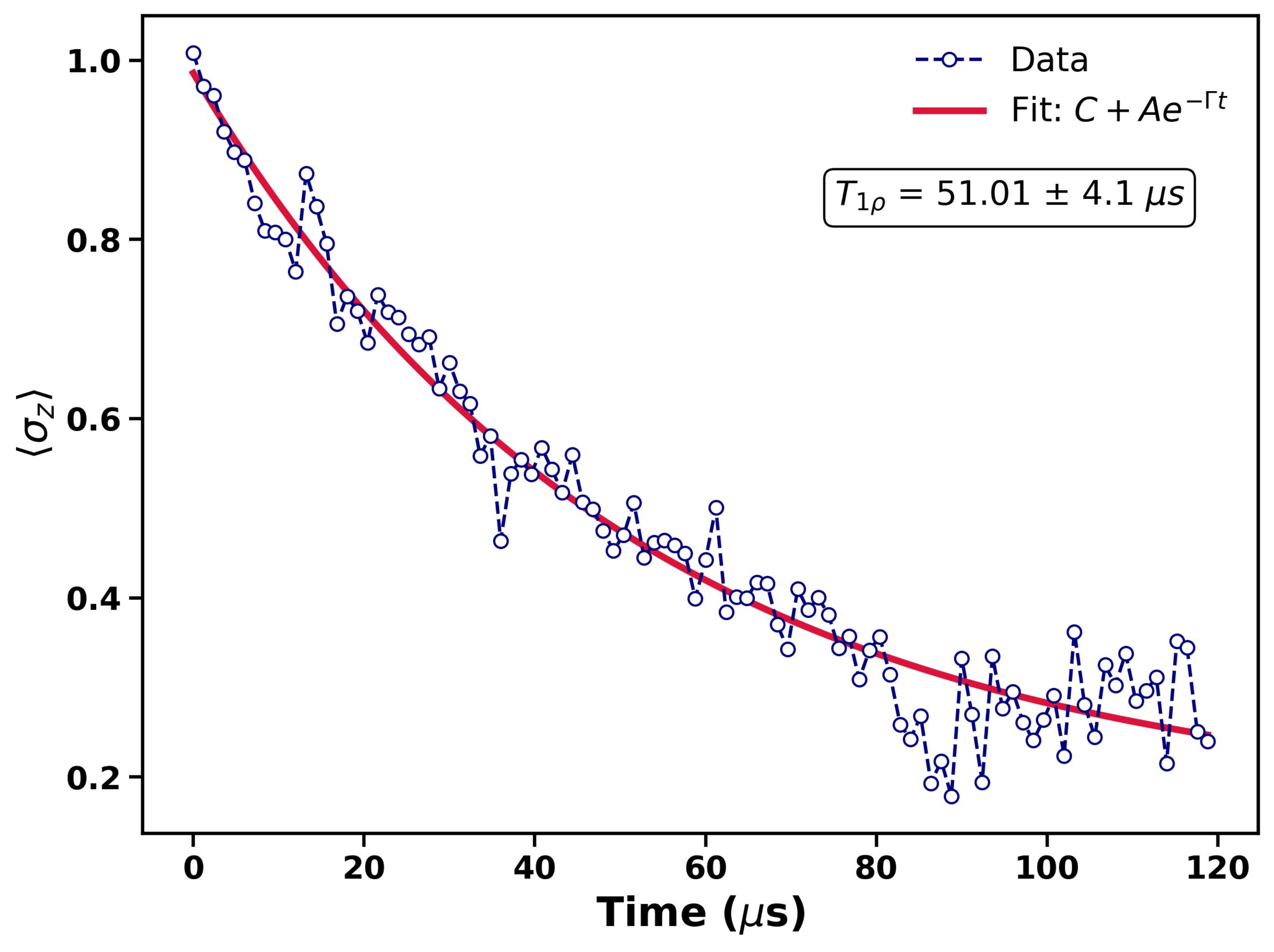}
    \caption{\justifying Measurement of the spin-locking relaxation time ($T_{1\rho}$) of the effective qubit under continuous $10~\mathrm{MHz}$ Rabi driving.}
    \label{fig_T1_rho_analysis}
\end{figure}

The tunability of the experimental system is indirectly verified by measuring qubit Rabi dynamics under various conditions and comparing the results with numerical simulations. The observed agreement confirms the controllability of the relevant system parameters. Notably, the measured dynamics emulate qubit dynamics of the quantum Rabi Hamiltonian in the ultrastrong-coupling regime, where the effective coupling strength between the effective qubit and the cavity mode is comparable to the effective qubit transition frequency \cite{Braumuller2017}.

Fig.~\ref{fig_various_rabi_dynamics} presents the measured effective qubit Rabi dynamics described by the effective quantum Rabi Hamiltonian in Eq.~(6) of the main text for three different effective qubit transition frequencies, $\Delta\Omega/2\pi = 0$, $0.07$, and $0.2~\mathrm{MHz}$. In particular, the Fig.~\ref{fig_various_rabi_dynamics}(a) corresponds to the resonant condition ($\Delta\Omega = 0$), where the cavity is continuously driven by a pair of double sideband tones detuned by $\pm10~\mathrm{MHz}$ from the cavity resonance, while the qubit is subjected to a resonant drive with a Rabi frequency of $10~\mathrm{MHz}$, thereby implementing the $i(\hat{a}-\hat{a}^\dagger)(\sigma_+ + \sigma_-)$ interaction in Eq.~(6) of the main text. Meanwhile, under this driving condition, the decay envelope of the resulting Rabi oscillation directly reflects the spin-locking coherence time of the effective qubit in the presence of the double sideband drive, yielding $T_{2\rho} = 17\mu\mathrm{s}$. (Experimentally, $T_{2\rho}$ was found to remain nearly unchanged even in the absence of the cavity double-sideband drive.) This independently measured coherence time was subsequently incorporated into the numerical master-equation simulations to accurately account for decoherence under the experimental conditions.

\section{Numerical master equation simulation} \label{Appendix D}

Compared with the previous theoretical model \cite{wn7t-pyrq}, the numerical master-equation simulations were extended to include the transmon anharmonicity up to the second excited state. In addition, the experimentally measured spin-locking relaxation and coherence times of the Rabi-driven effective qubit, $T_{1\rho}$ and $T_{2\rho}$, were incorporated as decoherence channels. Consequently, the simulations provide a more realistic description of the experimental system.

\begin{equation} \label{eq: System_H}
\begin{aligned}
    \frac{H(t)}{\hbar} &= \frac{H_q + H_c + H_I}{\hbar}\\
    &\frac{H_{q}}{\hbar} = \omega_q \hat{b}^\dagger \hat{b} - \frac{\alpha}{2}\hat{b}^\dagger \hat{b}^\dagger \hat{b}\hat{b} + \Omega_R(\hat{b}+\hat{b}^\dagger)\text{cos}(\omega_dt)\\
    &\frac{H_{c}}{\hbar} = \sum_{j=\{1,2\}}(\omega_{c_j}\hat{a}_j^\dagger \hat{a}_j + \epsilon_j(t)\hat{a}_j^\dagger+\epsilon_j(t)^*\hat{a}_j)\\
    &\frac{H_{I}}{\hbar} = \sum_{j=\{1,2\}}\frac{\chi_j}{2}(I-2\hat{b}^\dagger\hat{b}) \hat{a}_j^\dagger \hat{a}_j.
    \end{aligned}
\end{equation}
where
\begin{equation}
\begin{aligned}
    \epsilon_j(t)= \alpha_j\Omega_{SB}\text{cos}(\Omega_{SB}t)e^{-i\omega_{c_j}t}.
    \end{aligned}
\end{equation}
Here, $\hat{a}$ ($\hat{a}^\dagger$) denotes the annihilation (creation) operator of the cavity mode, and $\hat{b}$ ($\hat{b}^\dagger$) denotes the annihilation (creation) operator of the transmon.
The system dynamics were obtained by numerically solving the following master equation.
\begin{equation}
\begin{aligned}
\frac{d\rho}{dt}={}&-\frac{i}{\hbar}[H(t),\rho]
+\mathcal{D}\!\left[\sqrt{\frac{1}{2T_{1,\rho}}}\,\tilde{\sigma}_z\right](\rho) \\
&+\mathcal{D}\!\left[\sqrt{\frac{1}{2T_{2,\rho}^{*}}}\,\tilde{\sigma}_x\right](\rho)
+\sum_j\mathcal{D}\!\left[\sqrt{\frac{1}{T_{1c_j}}}\,\hat{a}_j\right](\rho),
\end{aligned}
\end{equation}
where $\tilde{\sigma}_z = |g\rangle\langle g| - |e\rangle\langle e|$, $\tilde{\sigma}_x = |g\rangle\langle e| + |e\rangle\langle g|$, and $\frac{1}{T_{2,\rho}^{*}} = \frac{1}{T_{2,\rho}} - \frac{1}{2T_{1,\rho}}$. $T_{1c_j}$ denotes the cavity photon lifetime of the $j$th cavity mode.

\section{Experimental setup and parameters} \label{Appendix F}

\begin{figure}
    \includegraphics[width = 0.48\textwidth]{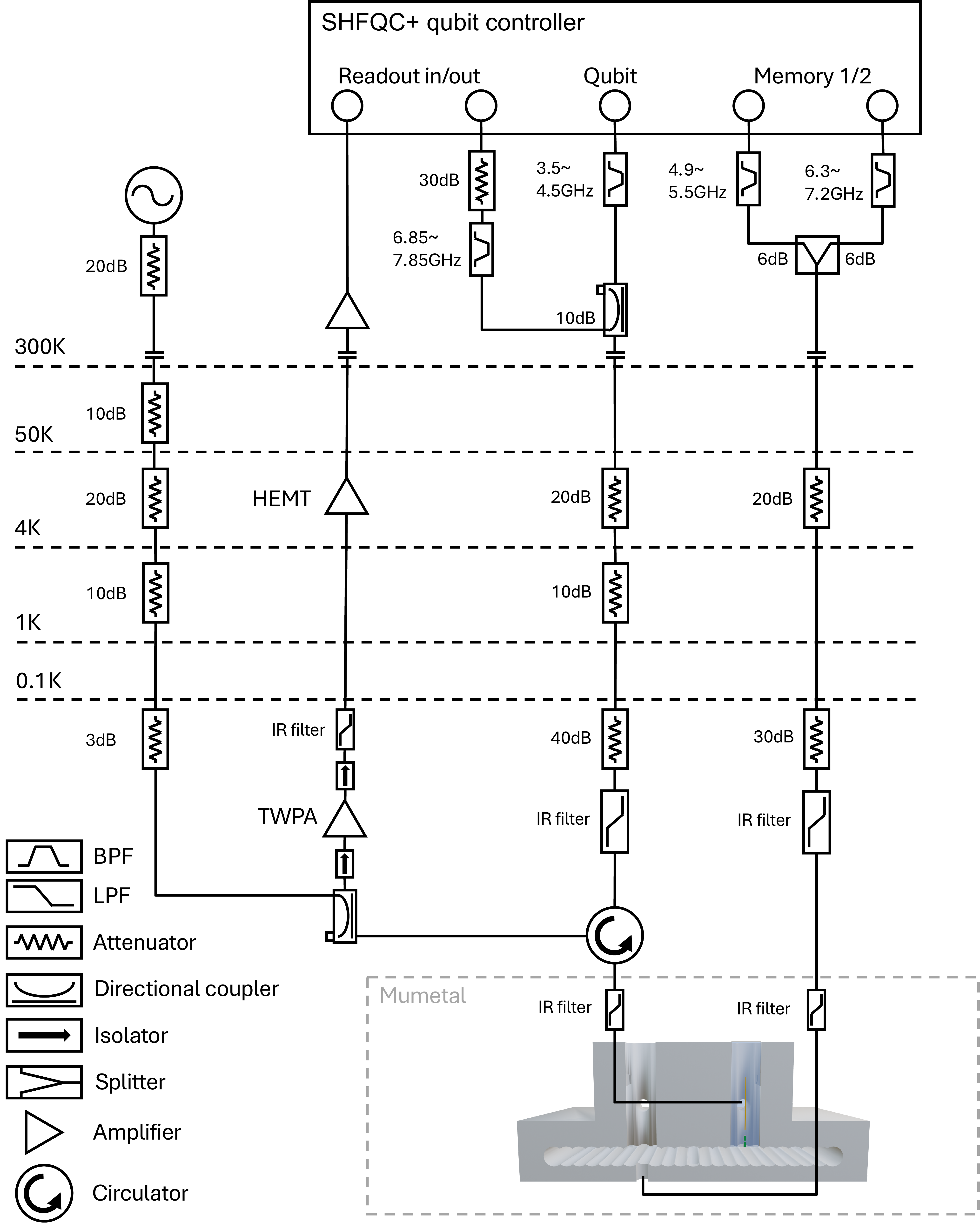}
    \caption{\justifying System connection of experimental setup}
    \label{fig_system_config}
\end{figure}

The experimental RF wiring configuration is shown in Fig.~\ref{fig_system_config}. And, the experimental parameters used in this work are summarized in Table~\ref{tab:parameters}.

\begin{table}
\caption{Experimental parameters. $T_{1\rho}$ and $T_{2\rho}$
are measured at a Rabi frequency of $10~\mathrm{MHz}$.}
\label{tab:parameters}
\centering
\begin{ruledtabular}
\begin{tabular}{lcc}
Quantity & Symbol & Value \\
\hline
1$_{st}$ cavity mode frequency & $\omega_{c_1}/2\pi$ & $5.1557$ GHz \\
2$_{nd}$ cavity mode frequency & $\omega_{c_2}/2\pi$ & $6.7996$ GHz \\
1$_{st}$ cavity mode relaxation rate & $\kappa_1/2\pi$ & $4.67$ kHz \\
2$_{nd}$ cavity mode relaxation rate                & $\kappa_2/2\pi$ & $15.09$ kHz \\
1$_{st}$ cavity mode dispersive shift & $\chi_1/2\pi$ & $-0.228$ MHz \\
2$_{nd}$ cavity mode dispersive shift                 & $\chi_2/2\pi$ & $-0.063$ MHz \\
Qubit frequency & $\omega_q/2\pi$ & $4.3064$ GHz \\
Charging energy & $E_c/h$ & $230$ MHz \\
Qubit coherence & $T_1,T_2,T_2^{\rm Echo}$ &
$35,25,35~\mu$s \\
Rotating-frame coherence & $T_{1\rho},T_{2\rho}$ &
$50,17~\mu$s \\
Readout frequency & $\omega_r/2\pi$ & $7.6642$ GHz \\
Readout relaxation rate & $\kappa_r/2\pi$ & $0.65$ MHz \\
\end{tabular}
\end{ruledtabular}
\end{table}

\section{Impact of experimental limitations} \label{Appendix E}

\begin{figure*}[p]
    \centering

    \begin{minipage}[t]{0.47\textwidth}
        \vspace{0pt}
        \centering
        \includegraphics[width=0.95\linewidth]
        {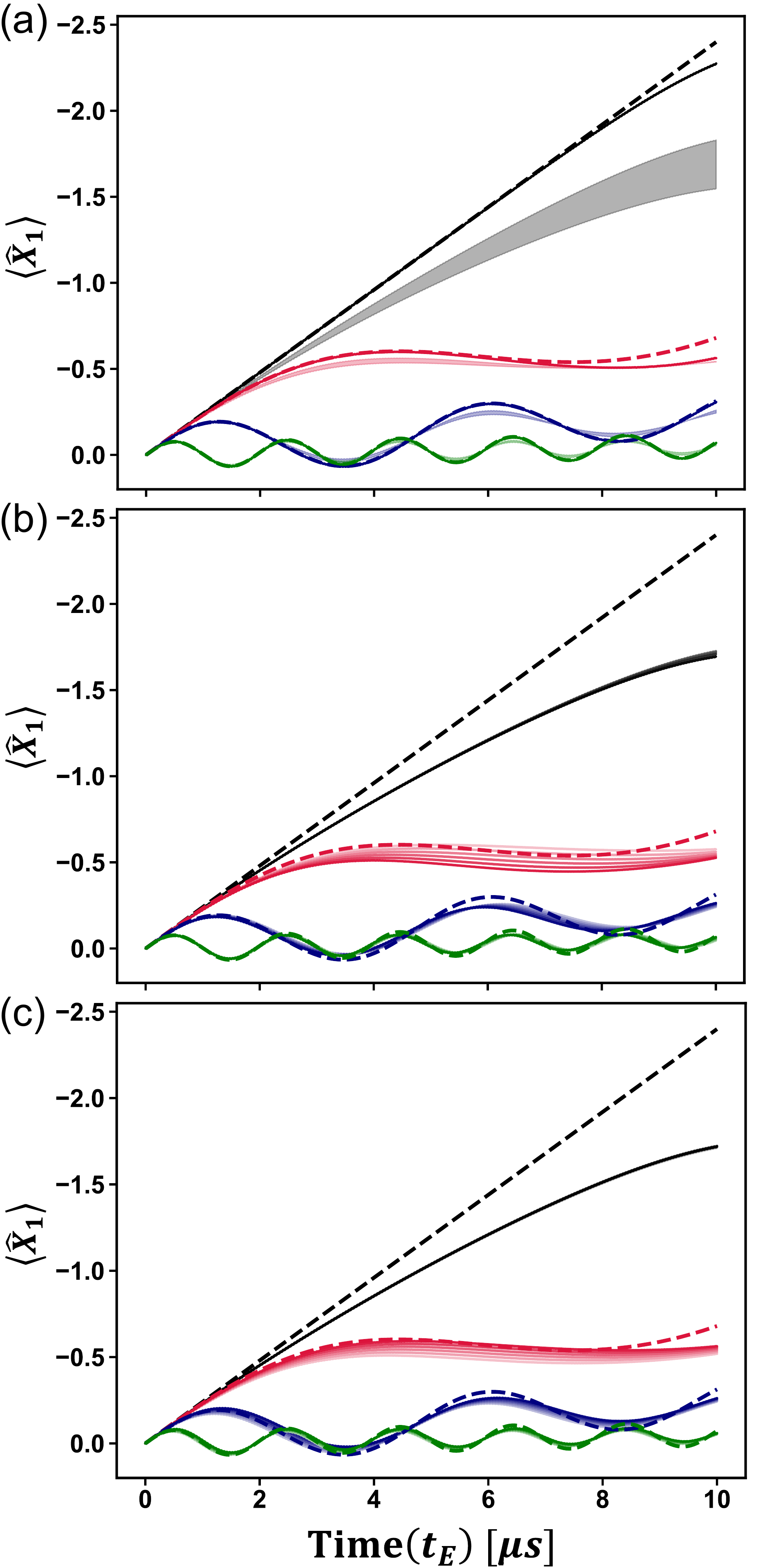}
        \caption{\justifying
        Numerical simulations illustrating the impact of realistic experimental limitations on the quantum simulation of one-dimensional Zitterbewegung for the effective masses considered in Fig.~(3) of the main text. (a) Effect of fluctuations in the spin-locking coherence time, $T_{2\rho}$. The shaded region corresponds to simulations with $T_{2\rho}$ varying from $12~\mu\mathrm{s}$ to $22~\mu\mathrm{s}$, while the solid line represents the ideal case without decoherence. (b) Effect of the finite precision of the Rabi-frequency calibration. Simulations are performed using the average decoherence rate and the nominal phase $\delta_j$, with the Rabi frequency varied within the experimentally achievable uncertainty of $\pm5~\mathrm{kHz}$. (c) Effect of the phase-calibration error. Simulations are performed using the average decoherence rate and the nominal Rabi frequency, while the phase $\delta_j$ is varied by $\pm5^\circ$. In both (b) and (c), the darkest (lightest) curve corresponds to the largest positive (negative) deviation from the nominal value.}
        \label{fig_appendix_g_1d}
    \end{minipage}
    \hfill
    \begin{minipage}[t]{0.47\textwidth}
        \vspace{0pt}
        \centering
        \includegraphics[width=0.9\linewidth]
        {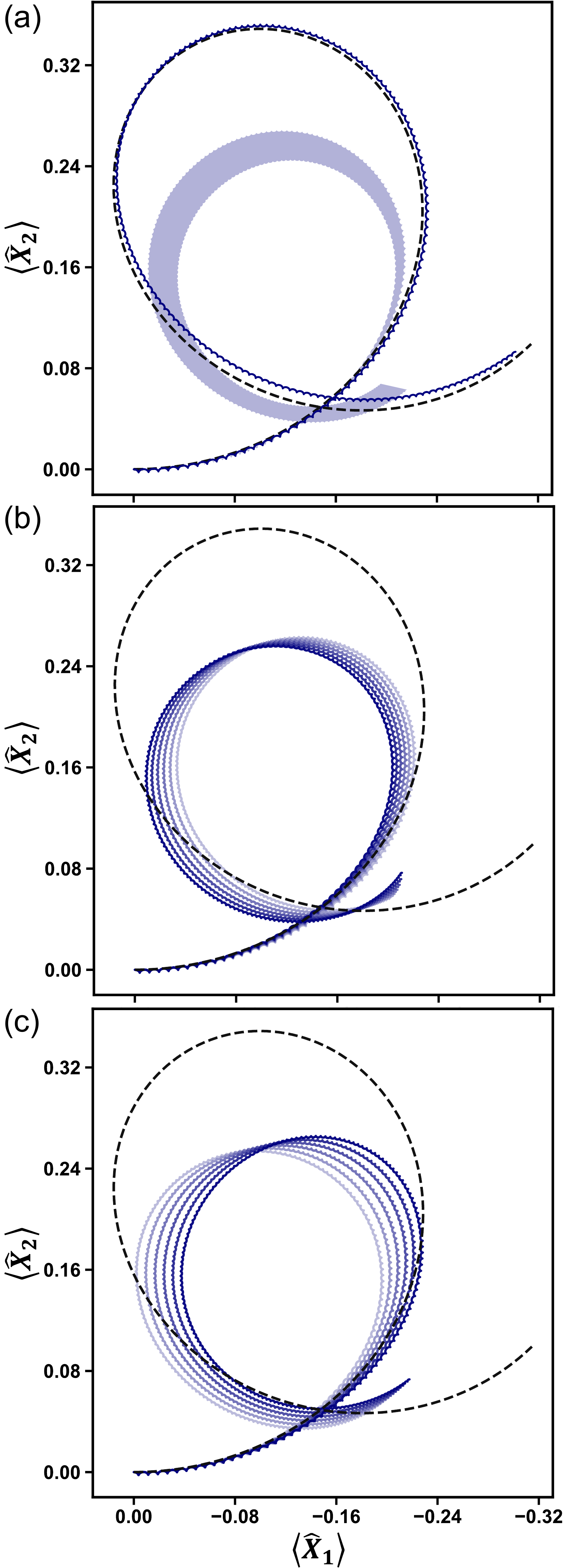}
        \caption{\justifying
        Numerical simulations illustrating the impact of realistic experimental limitations on the quantum simulation of two-dimensional Zitterbewegung for an effective mass of $\Delta\Omega/2\pi = 0.1~\mathrm{MHz}$. The panel descriptions are the same as those in Fig.~\ref{fig_appendix_g_1d}.}
        \label{fig_appendix_g_2d}
    \end{minipage}
\end{figure*}

To evaluate the impact of realistic experimental limitations on the quantum-simulated Zitterbewegung dynamics, we separately investigate the following four error sources: (i) temporal fluctuations of the spin-locking coherence time, $T_{2\rho}$, (ii) slow drifts of the Rabi frequency, (iii) the finite precision of the Rabi frequency calibration, and (iv) phase calibration errors of $\delta$. Their effects on the simulated one- and two-dimensional Zitterbewegung dynamics are evaluated separately through numerical simulations in Fig.~\ref{fig_appendix_g_1d} and Fig.~\ref{fig_appendix_g_2d}.

\clearpage

\end{document}